\documentclass[
reprint,
showpacs,
amsmath,amssymb,
aps,
prd,
]{revtex4-1}

\usepackage{color}
\usepackage[dvipsnames]{xcolor}
\usepackage{soul} 
\usepackage{graphicx}
\usepackage{dcolumn}
\usepackage{bm}
\usepackage{cancel}
\usepackage{mciteplus}
\usepackage{hyperref}

\begin{document}


\title{Restrictions from Lorentz invariance violation on cosmic ray propagation}


\author{H. Mart\'inez-Huerta}
 \email{hmartinez@fis.cinvestav.mx}
\affiliation{Departamento de F\'isica, Centro de Investigaci\'on y de Estudios Avanzados del I.P.N., Apdo. Post. 14-740, 07000, Ciudad de M\'exico, M\'exico}

\author{A. P\'erez-Lorenzana}%
 \email{aplorenz@fis.cinvestav.mx}
\affiliation{Departamento de F\'isica, Centro de Investigaci\'on y de Estudios Avanzados del I.P.N., Apdo. Post. 14-740, 07000, Ciudad de M\'exico, M\'exico}

\date{February, 2017}

\begin{abstract}
Lorentz Invariance Violation introduced as a generic modification to particle dispersion relations 
is used to study high energy cosmic ray attenuation processes. 
It is shown to reproduce the same physical effects for vacuum Cherenkov radiation,  as in some particular models with  spontaneous breaking of Lorentz symmetry. 
This  approximation is also  implemented for the study of photon decay in vacuum, where stringent limits to the violation scale are derived from the direct observation of very high energy cosmic ray photon events on gamma telescopes.
Photo production processes by cosmic ray primaries on photon background are also addressed, to show that Lorentz violation may turn off this attenuation process at energies above a well defined secondary threshold. 

\end{abstract}


\maketitle

\section{Introduction}\label{intro}

Lorentz symmetry stands as one of the cornerstones
of fundamental physics. Nonetheless, as for any other fundamental principle, exploring its limits of validity 
has been an important motivation for theoretical and expe\-ri\-mental research on the past \cite{NAMBU}.
Currently, supported mo\-dels with Lorentz Invariance Violation
(LIV) have renewed the interest of  the community due to the opportunity to observe or restrict new physics with the recent measurements from the Observatories at the highest ener\-gies.

Lorentz Invariance Violation is motivated as a possible consequence of beyond the Standard Model (SM) theories, such as Quantum Gravity or String Theory, just to mention some (see for instance Refs.~\cite{NAMBU,Bluhm, Pot,ALFARO,QG1,QG2,QG3,QG4,QG5,Gian}). However, the lack of observations of the LIV derived phenomena imposes limits on the validity scale of these models. 

Among the approaches used to introduce LIV on particle physics two stand out. One is generic and  it is not (necessarily) bounded to a particular model. The other one is based on the spontaneous breaking of the Lorentz symmetry. The former is implemented via  a general modification of the single particle dispersion relation, whereas the second mechanism introduces to the Standard Model a collection of LIV renormalizable operators that preserve energy-momentum and  microcausality, among other desired properties of the Standard Model.  These even have the advantage of being singlets of the Lorentz group  at coordinates level. This extension is named the minimal Lorentz-violating extension of the Standard Model (mSME)~\cite{SME}.

Both mechanisms, the mSME and the generic LIV, can lead to physics beyond Standard Model.  Nevertheless, the generic approach tends to converge quicker and simpler to phenomenology and it is not necessarily bound to a particular LIV model. Although, it can be compatible with the mSME,  they usually express their LIV parameters in different ways. 
Here, we propose to use the derived physics from the mSME as an {\it ansatz} to build a generic approach with the same physics, in order to contrast the results with experiments at the highest energetic window, that is, those observing cosmic and gamma rays. 
Astrophysical scenarios are an interesting place to test the possible signatures of LIV 
due to the high energies and very long distances they involve. The effects of LIV are expected to increase with energy for some energetic regime close to the Plank scale. Additionally, the very  long distances  can lead  to a significant LIV effect due to accumulative processes.
For the present study we have chosen to explore vacuum Cherenkov radiation  and photon decay for these purpose. Both processes are forbidden in the SM physics, but under LIV hypothesis they are permitted and can lead to extreme scenarios due to their implications on cosmic and gamma ray pro\-pagation. Once our generic approach had been validated, we will proceed to the exploration of other processes that are also relevant for cosmic ray propagation.

This article is organized as follows. On next subsection we will briefly present both, the spontaneous and the generic LIV mechanisms, that we will use along the paper.
On Sec.~\ref{VCR} we present the derived physics for va\-cuum Cherenkov radiation  from the mSME and  develop the generic approach for the same phenomena.  In Sec.~\ref{PD}, we focus on photon decay, using the LIV generic approach, to show that this phenomenon implies a very restrictive scenario for photon propagation. As a consequence of it, we extract limits for the LIV scale, $E_{QG}$,  to first and second order on the LIV correction, from the direct observation of very high energy photon events at H.E.S.S. and HEGRA cosmic gamma ray telescopes.
In addi\-tion, a general  analysis for photo production processes on the photon cosmic background, oriented to eva\-luate the effects of the generic LIV effects on the determination of the energy thresholds, is presented   
along section~\ref{REC}. Here, we also discuss the appearance of a se\-condary energy threshold, as a signature of LIV, above which photo production becomes forbidden.
Finally, a last  section is dedicated to summarize our results.

\subsection{Lorentz-violating mechanisms}\label{LIV}

\subsubsection{Generic LIV}

Generic LIV mechanism is perhaps the most commonly used on phenomenology, studies since it offers a clear idea on the derived physics from the LIV  hypo\-the\-sis  and it is not necessarily bound to any particular LIV-model. This mechanism converges to the introduction of an extra term in the dispersion relation of a single particle \cite{DIS1,DIS2,DIS3, JACOB, GUNTER-PH,GUNTER-PD, Stecker2009, Stecker2009NJ, GLASHOW_97, Gian}. Generically, this term can be motivated by the introduction of a not explicitly Lorentz invariant term at the free particle Lagrangian \cite{GLASHOW}. The extra term is restricted by a dimensionless coefficient that we call $\epsilon $.

To the lowest order, the  new dispersion relation for a free particle takes the form:
\begin{equation}\label{eq_dispersion}
	E^2-p^2 \pm \epsilon A^2  = m^2,
\end{equation}
where $(E,\mathbf{p})$ stand for the four-momenta  associated with a given particle of mass $m$, and we use the short hand notation $p=|\mathbf{p}|$ . $A$ can take the form of $E$ or $p$, however, for the ultra relativistic limit where $m\ll\{E, p\} $, any particular choice of $A$ will be equivalent. We will choose the sign of $\epsilon$ according to whether we correct energy or momentum,  leaving positive sign for energy.
For simplicity, we shall take $A= p$ in most of this article, but in section \ref{REC}, we will use $A= E$.
Asking that the LIV term can only be relevant at the highest energies, $\epsilon$ should be small, which guarantees that the physics at low energies remains Lorentz invariant. 

The most general modification to the dispersion relation would rather involve a general function of energy and momentum, such that one could write $E^2-p^2 + \epsilon(A)A^2 = m^2$. 
Since at low energies the contribution of $\epsilon$ should be negligible, one can use a
Taylor expansion and keep one term of order $n$ at once, assuming it as the leading term, 
in order to study the underlying physics. On this assumption 
the dispersion relation will become
\begin{equation}\label{eq_dispersion2}
	E^2-p^2 \pm \alpha_{n} A^{n+2} = m^2,
\end{equation}
where $\alpha_{n}=1/M^n = 1/(E_{QG}^{(n)})^n$ defines the energy scale associated to the LIV physics. $\alpha_n$ units are in $eV^{-n}$, hereafter.  $M$ is commonly  associated  with the energy scale of Quantum Gravity, $E_{QG}$,  which is expected to be close to the Planck scale, $E_{Pl}\approx 10^{19}GeV$. In literature, one can find upper limits to $E_{QG}^{(n)}$ by di\-ffe\-rent techniques,  even beyond the Planck scale \cite{GRB-LIV, HAWC-LIV, HEGRA-LIV, VERITAS-LIV,  PULSARS-LIV,   HESS-LIV, set, Example_n1, Example_n}.

Notice that Eq. (\ref{eq_dispersion2}) can be found from a particular LIV model. For instance, Coleman and Glashow formalism is compatible with the special case $n=0$ \cite{GLASHOW,GLASHOW_97}, likewise, for $n=1$ see for instance \cite{Example_n1} and for any $n$ see Ref.~\cite{Example_n}.

\subsubsection{mSME}

The mSME provides  a more specific theo\-retical ground to study the resulting physics from the introduction of the LIV hypothesis, throughout the spontaneous  violation of Lorentz symmetry \cite{NAMBU, Bluhm}. This me\-cha\-nism  adds to the Standard Model (SM) all observer-scalar terms that are products of SM and gravitational fields 
which are not Lorentz invariant, and generically require couplings that explicitly carry Lorentz indices. 
The minimum set of such operators that maintain gauge invariance and power-counting renormalizability conform the so called  mSME and was proposed by Colladay and Kostelecky~\cite{SME}.
A first classification of these new terms corresponds to the own sectors of SM, such as the $\mathcal{L}_{lepton}$ or $\mathcal{L}_{gauge}$, for lepton or gauge sectors respectively.  Additionally, such operators can be classified into those that break CPT symmetry and those that do not. They are named \textit{CPT-odd }and \textit{CPT-even} respectively.

In the present work, we have chosen the photon sector with the following additional contributions to the standard Lagrangian density,
\begin{equation}\label{eq_ph_even}
	\mathcal{L}^{CPT-even}_{photon} = 
	-\dfrac{1}{4}(k_F)_{\rho\lambda\mu\nu}F^{\rho\lambda}F^{\mu\nu},
\end{equation}
and
\begin{equation}\label{eq_ph_odd}
	\mathcal{L}^{CPT-odd}_{photon} = 
	\dfrac{1}{2}(k_{AF})^{\rho}\epsilon_{\rho\lambda\mu\nu}A^{\lambda}F^{\mu\nu},
\end{equation}
with the gauge field $A_\mu$ and the field strength tensor $F_{\mu\nu}=\partial_\mu A_\nu - \partial_\nu A_\mu$. In above $\epsilon_{\mu\nu \rho\lambda}$ is the antisymmetric Levi-Civita symbol.
The coefficients $k_F$ and $k_{AF}$ in Eqs. (\ref{eq_ph_even}) and (\ref{eq_ph_odd}) are the ones that break particle Lorentz Invariance. Besides, $k_F$ is dimensionless and $k_{AF}$ has mass dimension one. 
On the analysis regarding this sector, it is usual not to assume the simultaneous presence of LIV corrections for fermions and we are going to do so in our own study along the following sections.

The {\it CPT-even}  term can be studied separately as in the so called modified Maxwell theory (modM \cite{TWO-SIDE}),
\begin{equation}\label{modM}
	\mathcal{L}_{modM}= -\dfrac{1}{4}F_{\mu\nu}F^{\mu\nu} 
	-\dfrac{1}{4}\kappa^{\mu\nu}_{\ \ \rho\lambda}F_{\mu\nu} F^{\rho\lambda}.
\end{equation}
$\kappa^{\mu\nu}_{\ \ \rho\lambda} $ is a dimensionless fourth rank tensor, with  256 components, but it its usually assumed to have vanishing double trace, $\kappa^{\mu\nu}_{\ \ \mu\nu}=0$,
  and demanded to obey the same symmetries which hold for the Riemann curvature tensor,
\begin{equation} \label{riemann}
	\kappa _{\mu\nu \rho\lambda}=-\kappa_{\nu\mu \rho\lambda}=\kappa_{\nu\mu \lambda\rho}, \ \ 
	\kappa_{\mu\nu \rho\lambda}=\kappa_{\rho\lambda \mu\nu},
\end{equation} 
such that only 19 independent components remain~\cite{3-LEVEL}.
However, the following {\it ansatz}, used in Ref.~\cite{TWO-SIDE}, reduces the number of independent parameters from 19 to 1. 
\begin{equation} \label{a_nb}
	\kappa^{\mu\nu \rho\lambda}=\dfrac{1}{2} 
	(\eta^{\mu\rho}\tilde{\kappa}^{\nu\lambda}
	-\eta^{\mu\lambda}\tilde{\kappa}^{\nu\rho} +\eta^{\nu\lambda}\tilde{\kappa}^{\mu\rho} 
	-\eta^{\nu\rho}\tilde{\kappa}^{\mu\lambda} ),
\end{equation} 
and
\begin{equation} \label{a_i}
	\tilde{\kappa}^{\mu\nu} = \dfrac{3}{2} \tilde{\kappa}_{tr} \  \text{diag}(1,1/3,1/3,1/3),
\end{equation} 
where Eq. (\ref{a_nb}) restricts the theory to the nonbirefringent sector and Eq. (\ref{a_i}) does it to the isotropic sector.

On the other hand, the CPT-{\it odd} term is stu\-died in the spacelike Maxwell-Chern-Simons theory (MCS \cite{VCR-MCS}), with the following Lagrangian density,
\begin{equation}\label{MCS}
	\mathcal{L}_{MCS}= -\dfrac{1}{4}F_{\mu\nu}F^{\mu\nu} 
	+\dfrac{1}{4}M_{CS}\epsilon_{\mu\nu \rho\lambda}\xi^\mu A^\nu F^{\rho\lambda},
\end{equation}
where  $M_{CS}\sim 1/L_0 \approx 2 \times 10^{-33} eV$ is the Chern-Simons mass scale~\cite{CS-scale,VCR-MCS} and $\xi_\mu=(0,0,0,1)$ is the fixed space-like normalized Chern-Simons vector.

It is noteworthy that in Eqs. (\ref{eq_dispersion2}), (\ref{eq_ph_even}) and (\ref{eq_ph_odd}) the involved coefficients are not necessarily the same and so, numerically, 
\begin{equation}
	k_F \neq k_{AF} \neq \alpha_n,
\end{equation}
despite that all of them parametrize the LIV hypothesis.  Hence, the va\-lues and limits that can be derived for them shall be different too. Therefore, for comparison proposes with the LIV generic approach, we will use the limit va\-lues given in the literature for the involved parameters of modM and MCS models.

\section{Vacuum Cherenkov Radiation}\label{VCR}

\subsection{Emission Rate}

When electrically charged and massive particles  ruled by a Lorentz invariant physics are added to the LIV mo\-dels presented above, the new terms allow photon emission in vacuum to happen. This process is  commonly called vacuum Cherenkov radiation  (See for instance \cite{GLASHOW,GLASHOW_97,CROSS, TWO-SIDE, VCR-MCS}). In a general scenario, the derived physics will be spin dependent, and thus, either particles with spin 0 or $1/2$ will be able to emit photons. To further proceed with our analysis for the generic implementation of LIV and the latter comparison with the outcomes of the modM and  MCS models,   
we will restrict our discussion  to spin $1/2$ particles.  In what follows, let us first summarize the results for these models.

\begin{figure}[!ht]
\centering
	\includegraphics[width=.2\textwidth]{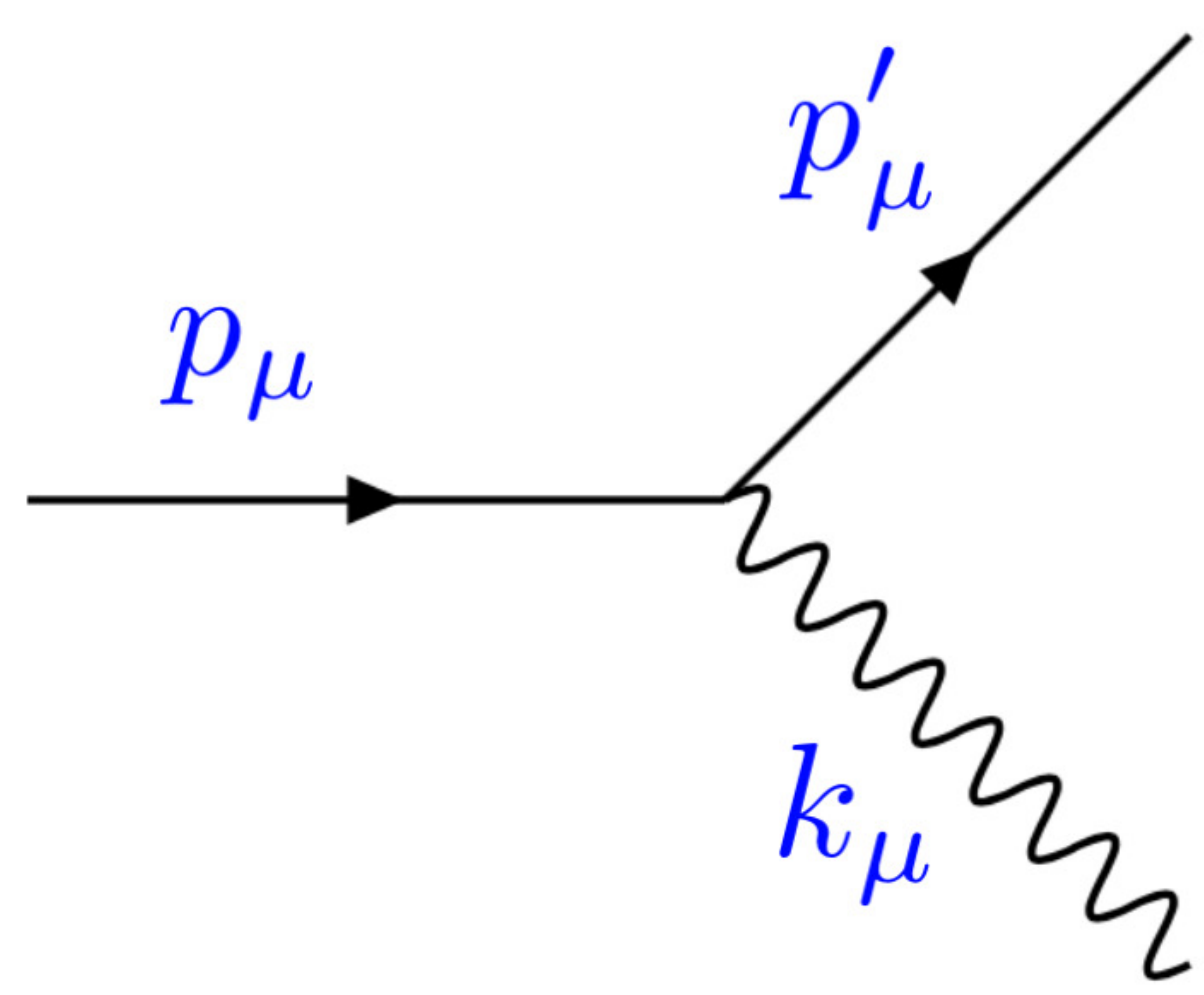}
		\caption{Spontaneous photon emission, or vacuum Cherenkov radiation, Feynman diagram.}\label{VCR_vertex}
\end{figure}

The process of interest, $l_{p} \rightarrow l_{p'} \ \gamma_k$,  is depicted in Fig.~\ref{VCR_vertex}, where 
the vacuum Cherenkov radiation may co\-rres\-pond in general to leptons, protons and heavy nuclei. The subscript denotes the momentum notation. Hereafter, $(\omega, \mathbf{k})$ will stand for the photon four-momenta components. 
Also, $(E,\mathbf{p})$ will represent  the initial four-momentum of the spin $1/2$  charged particle involved in the process, while $(E',{\mathbf{p}}')$ will stand for the final four-momentum components.
Following Refs.~\cite{TWO-SIDE, VCR-MCS}, the dispersion relations for a single photon from modM and MCS models are written as
\begin{equation}
	\omega_{modM} (\mathbf{k})=\sqrt{\dfrac{1-\widetilde{\kappa}_{tr}}{1+\widetilde{\kappa}_{tr}}}~k, ~~~
	k=|\mathbf{k}|,
\end{equation}
\begin{equation}
	\omega_{MCS \ \pm} (\mathbf{k})^2 = k^2 \pm M_{CS}\sqrt{k^2 \cos^2\phi + \frac{M_{CS}^2}{4}} + \frac{M_{CS}^2}{2},
\end{equation}
where $\tilde{\kappa}_{tr}$ is the restricted coefficient from Eq.~(\ref{a_i}). 
In the MCS expression, $\phi$ is the angle between the wave vector $\mathbf{k}$ and the space component of the Chern-Simons vector, $\mathbf\xi$. Notice that this model  is not rotationally invariant, so energy-momentum will be conserved but angular momentum shall not. The subscript $\pm$ denotes two polarization modes.

The reported emission rates for these models are
\begin{equation}\label{eq_modM_G}
{\small
\begin{aligned}
	\Gamma_{modM}  =  & \frac{\frac{e^2}{4\pi}}{3\tilde{\kappa}_{tr}^2\sqrt{1-\widetilde{\kappa}_{tr}^2}} \frac{1}{E} 
	 	\left(  \sqrt{\frac{1-\widetilde{\kappa}_{tr}}{1+\widetilde{\kappa}_{tr}}}  \frac{E}{\sqrt{E^2-m^2}}-1 \right) \\
 		\times \bigg[ & \sqrt{\frac{1-\widetilde{\kappa}_{tr}}{1+\widetilde{\kappa}_{tr}}}				  (3\widetilde{\kappa}_{tr}^2+5\widetilde{\kappa}_{tr}+2)E\sqrt{E^2-m^2}  \\
	 	& - (3\tilde{\kappa}_{tr}^2+3\tilde{\kappa}_{tr}+2)E^2 + (3\tilde{\kappa}_{tr}^2+4\tilde{\kappa}_{tr}+1)m^2  \Big],
\end{aligned}
}
\end{equation}
and
\begin{equation}
{\small
	\begin{aligned}
	\Gamma_{MSC} = & \frac{\frac{e^2}{4\pi} M_{CS}}{16E((\mathbf{p}\cdot\xi)^2 + m^2)^{1/2}}	
	\Bigg\{ \Big[4\Big(m^2+2(\mathbf{p}\cdot\xi)^2\Big)-M_{CS}^2 \Big] \\
	& \times  arcsinh\left(\frac{2k_{max}}{M_{CS}}\right) + 2(2|\mathbf{p}\cdot\xi| - k_{max}) M_{CS} \\
	& -4 (|\mathbf{p}\cdot\xi|-k_{max}) \sqrt{M_{CS}^2 + 4k_{max}^2}  - 8 m^2\dfrac{k_{max}}{M_{CS}}\Bigg\},
 	\end{aligned}
}
\end{equation}
where
\begin{equation}
{\small
	\begin{aligned}
	k_{max} = \frac{2M_{CS}|\mathbf{p}\cdot\xi|(M_{CS}+2\sqrt{(\mathbf{p}\cdot\xi)^2 + m^2})}{M_{CS}^2+4m^2 + 		4M_{CS}\sqrt{(\mathbf{p}\cdot\xi)^2 + m^2}},
	 \end{aligned}
}
\end{equation}
is the maximum magnitude for the photon momentum component for MCS emission rate.

Next, let us elaborate the corresponding analysis for the generic case. So far, there is not a clear path to derive the emission rate from the LIV generic approach. However, we will follow the generalities from the two previous models to construct it. 
Starting with Eq. (\ref{eq_dispersion2}), the dispersion relation for  photons can be written as
\begin{equation}\label{eq_omega}
	\omega^2 = k^2 ( 1 + \alpha_n k^n ),
\end{equation}
where $\alpha_n$ is the LIV coefficient of order $n$.
We shall also neglect, as in previous models, any LIV effect in the corresponding fermion dispersion relations.
Then, we will develop a generic correction to the emission rate from the standard LI theory, given by:
\begin{equation}\label{eq_deltaVCR}
	\begin{aligned}
	d\Gamma = & \dfrac{s}{2} \dfrac{1}{E(p)} |\mathcal{M}|^2 \dfrac{d^3p'}{(2\pi)^3 2 E'(p')} \dfrac{d^3 k}{(2\pi)^3 2 \omega (k,n,\alpha_n)} \\
	& \times (2\pi)^4\delta^{(4)}( p-p'-k )\Theta(k),
	\end{aligned}
\end{equation}
where  we  have inserted a Heaviside function,  $\Theta(k)$, in order to preserve only physical solutions for the emitted photon momentum. Also, 
$s=\frac{1}{j!}$ is an statistical factor, where $j$ counts one per each group of identical particles in the final state. $|\mathcal{M}|^2$ is the squared amplitude probability of the process, computed from the standard QED rules but making use of the correction in the photon four-momenta. 
After the calculations we found that
\begin{equation}\label{eq_M}
	|\mathcal{M}|^2 = e^2 | 4m^2-\alpha_n k^{n+2}|, 
\end{equation}
where orthogonality relation is assumed for  photon at first approximation, $\epsilon_\mu \bar{\epsilon}_\nu \approx g_{\mu\nu}$. 
For a more detailed calculation  see  Ref. \cite{Proc2}.
The delta function in  Eq. (\ref{eq_deltaVCR}) encodes the conservation of energy-momentum in the system. Using  Eq. (\ref{eq_omega}) 
and keeping only first order in LIV $\alpha_{n}$ terms, we made the integration over $\mathbf{p'}$ in Eq. (\ref{eq_deltaVCR})  and  write the remaining  delta function as:
\begin{equation}{\small
	\begin{aligned}
	& \delta^{(0)}(g(p,k,m,n,\alpha,\theta))  = \\
	& \delta^{(0)}(\sqrt{p^2+m^2}-\sqrt{(\mathbf{p}
	    -\mathbf{k})^2+m^2}
	    -k\sqrt{1+\alpha_nk^n}) 
	\end{aligned}, }
\end{equation}
where $\theta$ is the emission angle and $g$ is a polynomial function of order $n$ in $k$.
To solve the integration in $k$ without the rest frame, that is, only in the observer frame, we used the following well known property of the delta function,
\begin{equation}\label{eq_delta}
	\delta(g(x))=\sum_i \dfrac{\delta(x-x_i)}{|g'(x_i)|},
\end{equation}
where $x_i$ are the $g(k_i)$ roots, that we can found from the following expression,
\begin{equation}\label{eq_VCR_g}
{\small
	\begin{aligned}
	\left(p^2 \cos^2 \theta - (p^2+m^2)\right) k - & (p^2+m^2) \alpha_n k^{n+1}  \\
	& + p \cos \theta \alpha_n k^{n+2}=0.
	\end{aligned}
	}
\end{equation}
Each solution of Eq. (\ref{eq_VCR_g}) implies that the process will satisfy energy-momentum conservation. 
Note that if $\alpha_n=0$ then $k=0$, since $p \cos \theta \neq \sqrt{p^2+m^2}$, and consequently, the process is prohibited for LI scenarios. In other words, the photon emission process in vacuum, or vacuum Cherenkov radiation, is  permitted as far as $\alpha_n$ is present.

For $n=1$, we  found that Eq. (\ref{eq_VCR_g}) has three different solutions. That is, in the generic LIV scenario energy and momentum are preserved in at least three different situations, given by the standard case where
\begin{equation}
k_0=0;
\end{equation}
and when
\begin{equation}\label{eq_kplus}
\small
	\begin{aligned}
	k_\pm = & \frac{1}{2p\cos\theta} \left( E^2   \pm \sqrt{E^4 +  \frac{4p\cos\theta}{\alpha_{1}}(E^2 - p^2\cos^2\theta)}\right).
	\end{aligned}
\end{equation}
In what follows we will only use the mode $k_+$, since $k_0$ means that there is none emission and  $k_-$ will be ruled out by the Heaviside function in Eq. (\ref{eq_deltaVCR}), besides it is nonphysical.

Finally, one can use the expressions in Eqs.~(\ref{eq_omega}), (\ref{eq_M}) and that for $k_+$ in Eq.~(\ref{eq_kplus}), to perform the integral over $k$. 
Hence, the emission rate from the LIV generic approach at first order in $\alpha_{n}$ (LIVgen1) will be expressed as,
\begin{equation}\label{eq_deltaVCR_VF}
{\small
	\begin{aligned}
	& \Gamma
	 =   \dfrac{e^2}{4\pi} \dfrac{1}{4E} 
 	\int_0^{\theta_{max}} \frac{|4m^2 - \alpha_1k_+^3| }{\omega(k_+,\alpha_{1} )} \\
	   & \times \frac{k \sin\theta d\theta}{|-k_+ + p \cos\theta - \frac{(1+\frac{3}{2}\alpha_1k_+)}{\sqrt{1+\alpha_{(1)}k_+}} \sqrt{k_+^2 + E^2-2k_+p\cos\theta}|},
	\end{aligned}
}
\end{equation}

\begin{figure}[!ht]
	\centering
		\includegraphics[width=.48\textwidth]{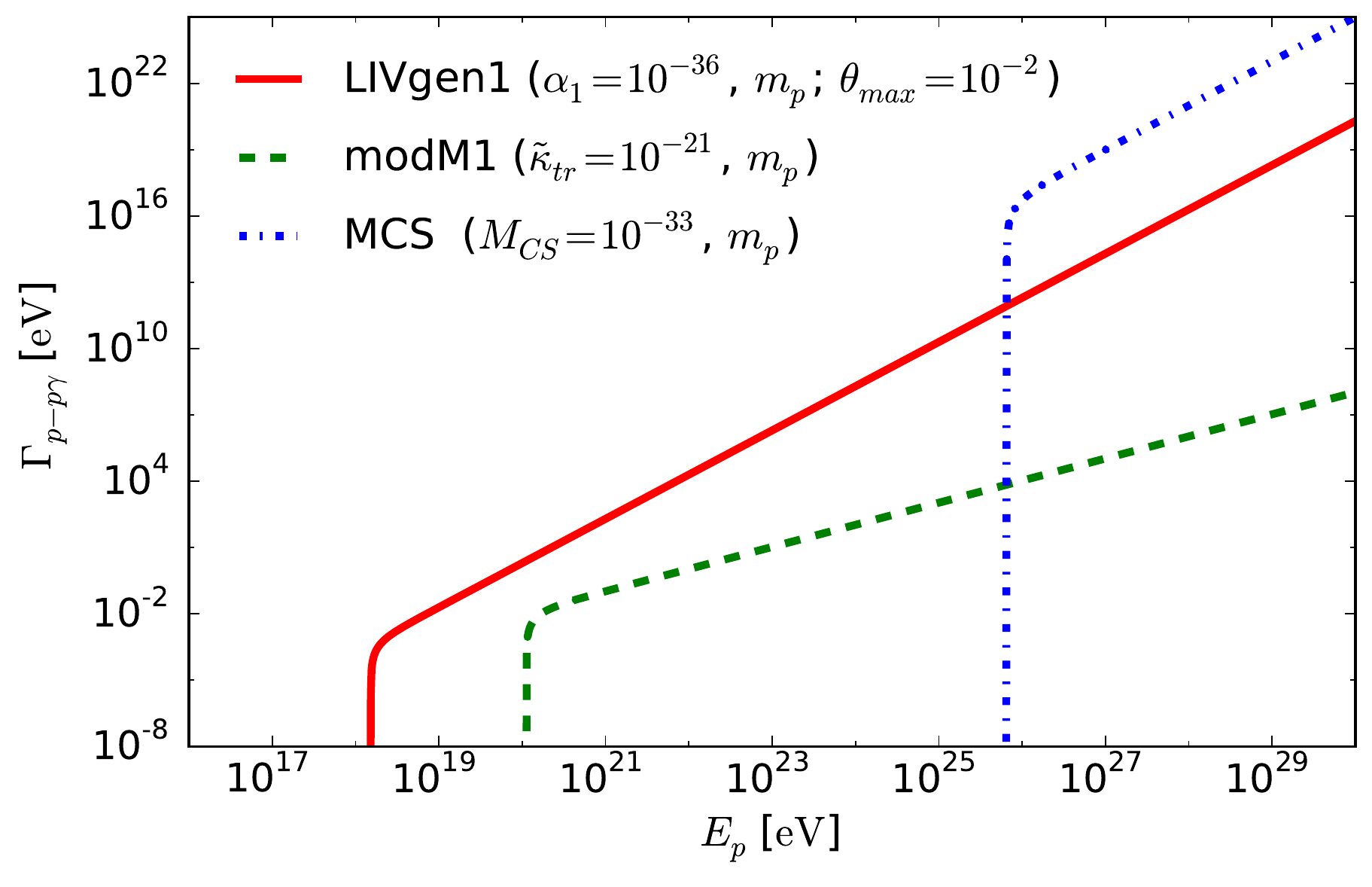}
		\caption{Comparison of emission rates for vacuum Cherenkov radiation for a proton. Free LIV parameters were chosen as indicated}\label{plot_VCR-gamma}
\end{figure}

In figure~\ref{plot_VCR-gamma}, we show the emission rate for the three di\-ffe\-rent models for proton mass, $m=m_{p}$.  
We have plotted, in the (green) dashed line, the modM emission rate at the first order from the expansion on $E$ and $\tilde{\kappa}_{00}$ from Eq. (\ref{eq_modM_G}) and setting $\tilde{\kappa}_{00}=10^{-21}$,  which is compatible with the limits reported in Ref. \cite{HEGRA-LIV}~. From a phenomenological point of view, the rate has four main characteristics. First, it has a growing behavior with the charged particle energy. Second, it has  a drop ending threshold for the model,  inversely proportional to the LIV coefficient. This threshold protects the LI physics at energies below it.
Third, for a given energy above the threshold, the emission rate decreases for smaller values in the LIV parameter. And fourth, it has a sensitive behaviour with the mass and charge of the emitting particle. The last two characteristics will be discussed in the next section.
The MCS emission rate, in dash-dotted (blue) line in the same figure, shows the same behavior as modM but the  derived threshold is higher in energy than the previous one, since it is highly constrained from the $M_{CS}$.
Finally, we present in a (red) continuous line the emission rate from the LIVgen1 model for very small $\theta$ and with a phenomenological threshold given by $E \gtrsim \left(\frac{4m^2}{\alpha_1}\right)^{1/3}$.
Hence, the generic approach  reproduces the main features of the previous two models.
We have found that the case $n=0$ has a similar behaviour \cite{Proc3}.

\subsubsection{Sensitivity to UHECR mass composition}

\begin{figure}[!ht]
	\centering
	\includegraphics[width=.48\textwidth]{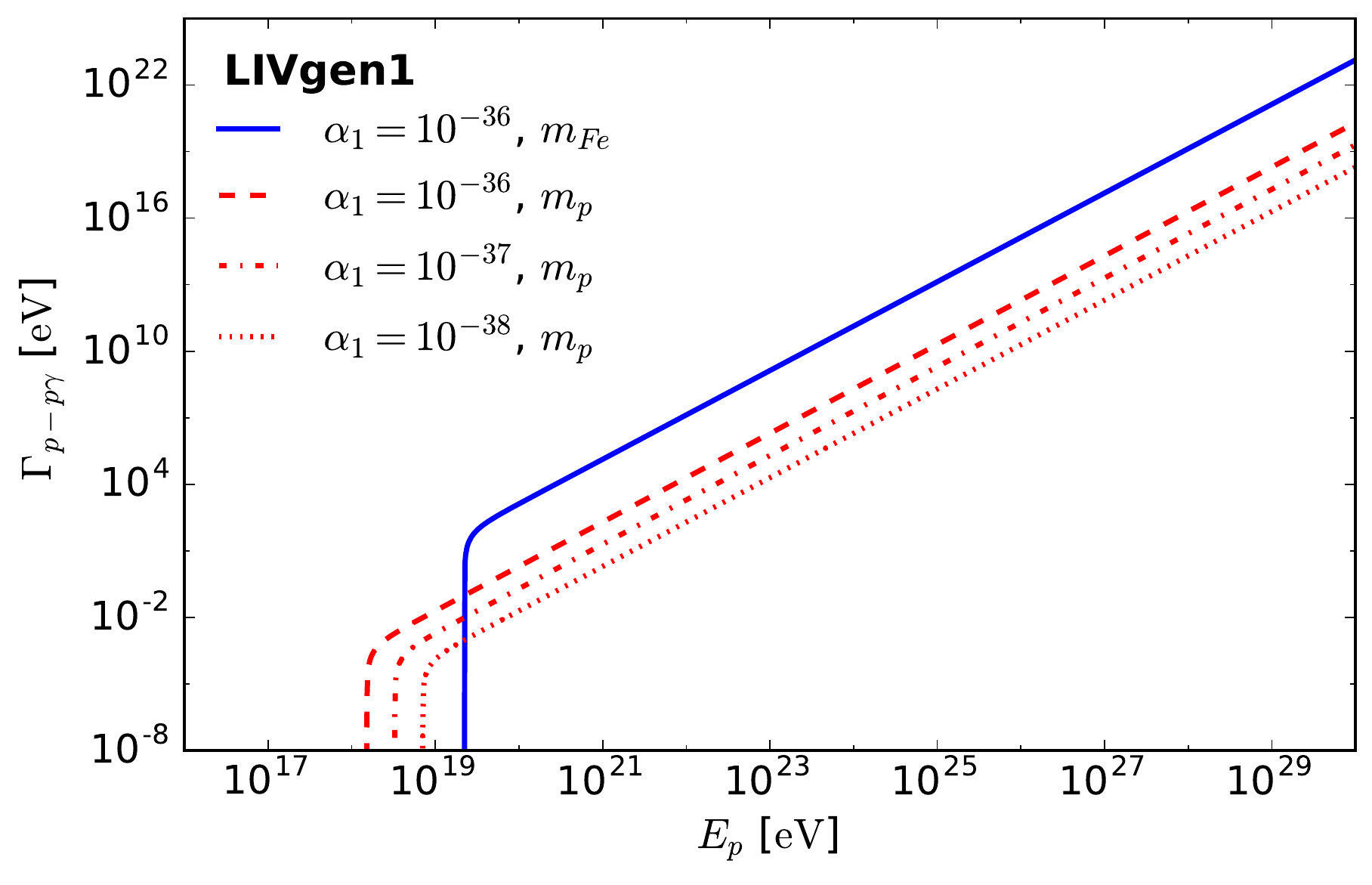}
	\caption{ Emission rates for vacuum Cherenkov radiation from LIV generic approach at first order correction, with $n=1$ and $\theta_{max}=10^{-2}$. Corresponding rates for $m_{Fe}$, $m_p$ and different $\alpha_{1}$ are as indicated. Both rates grow with energy, but lighter particles have a lower emission rate. The energy threshold is sensitive to particle composition.}\label{plot_VCR-massLIV}
\end{figure}

\begin{figure}[!ht]
\centering
	\includegraphics[width=.48\textwidth]{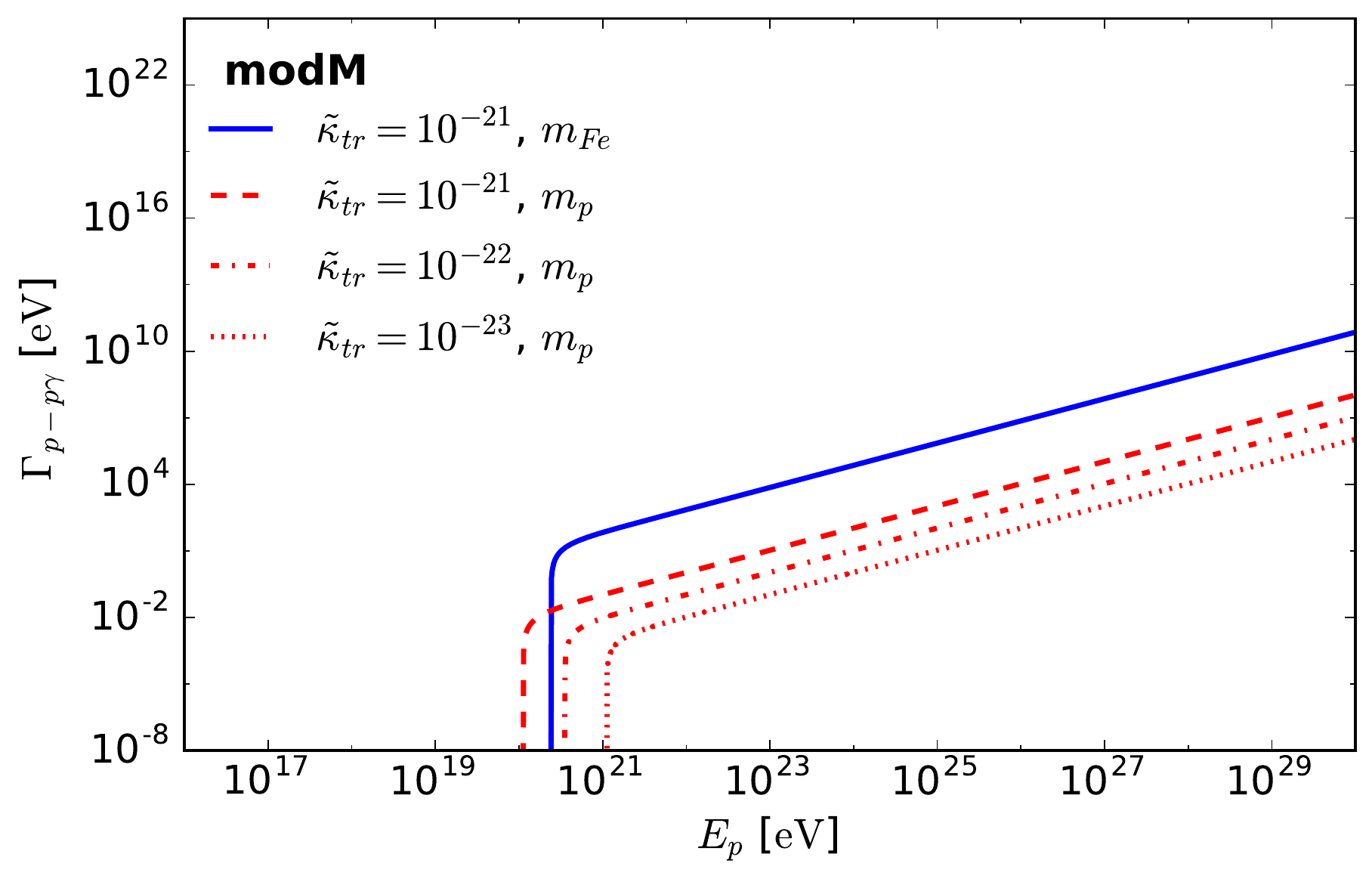}
	\caption{Emission rates for vacuum Cherenkov radiation from modified Maxwell theory. Corresponding rates for $m_{Fe}$, $m_p$  and different $\widetilde{\kappa}_{tr}$ are as indicated. Both rates grow with energy, but lighter particles have a lower emission rate. The energy threshold is sensitive to particle composition.}\label{plot_VCR-massMM}
\end{figure}

Considering the application of these models to the phenomenon of cosmic rays, we have chosen  $m=m_p$ and $m=m_{Fe}$ as  representative masses for a light and a heavy component of the cosmic ray flux, respectively. It is estimated that the composition of ultra high energy cosmic rays (UHECR) is mixed and varies between these two limits~\cite{MIXED}.  
To appreciate the differences between the components, in Figs. \ref{plot_VCR-massLIV} and \ref{plot_VCR-massMM}, the  behavior of the $\Gamma$ functions from LIVgen1 and modM for proton and iron are compared. 
From  them, it can be seen that the effect of vacuum Cherenkov emission is sensitive to the composition of cosmic rays. 
\begin{figure}[!ht]
	\centering
	\includegraphics[width=.49\textwidth]{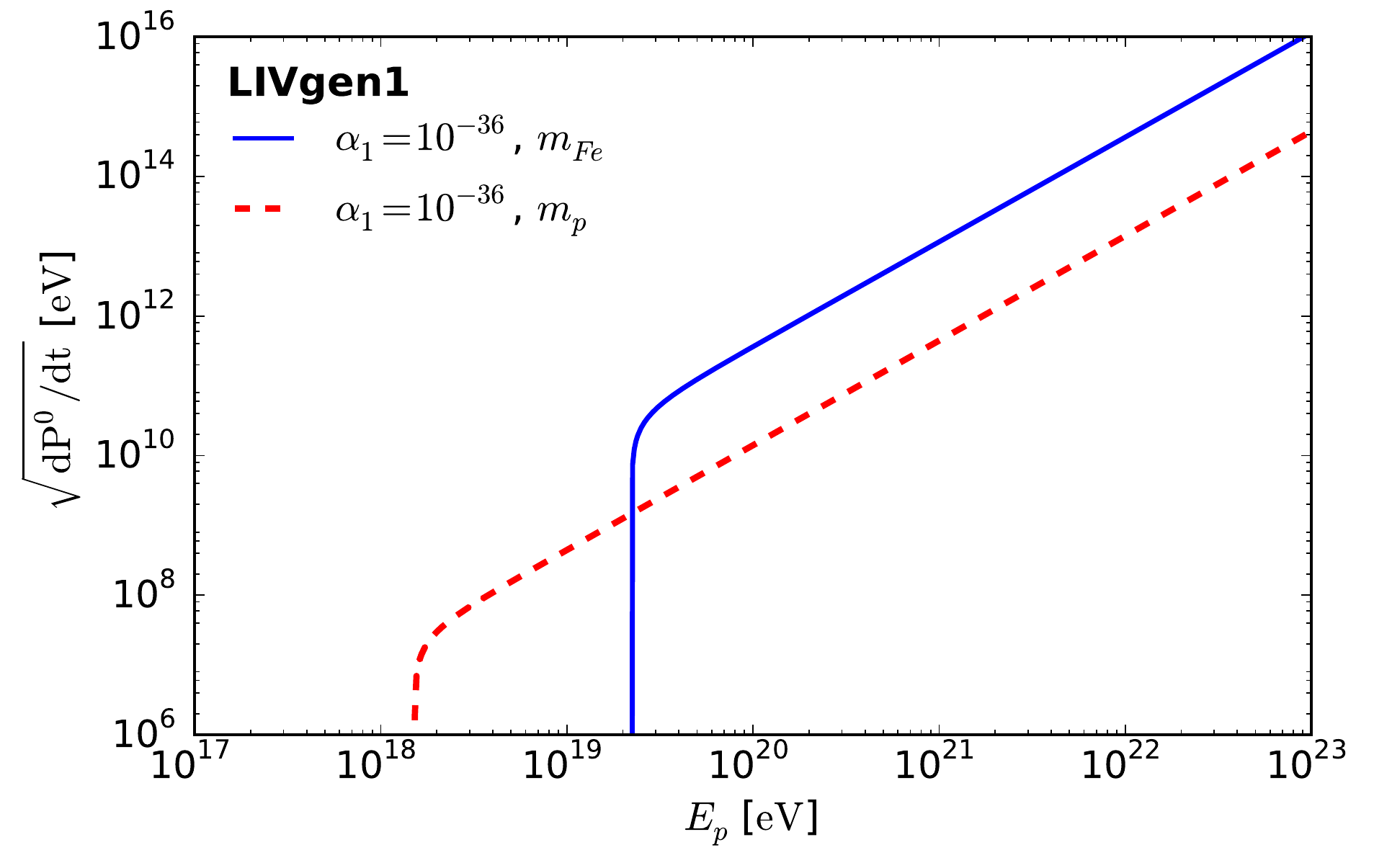}
	\caption{Radiated energy  for vacuum Cherenkov radiation from LIV generic approach at first order correction, with $n=1$ and $\theta_{max}=10^{-2}$. Corresponding curves for $m_{Fe}$ and $m_p$ are as indicated.}\label{plot_VCR-RE-LIV}
\end{figure}
In Figs. \ref{plot_VCR-massLIV} and \ref{plot_VCR-massMM}, we also  stand out the emission rate behaviour for smaller values in the LIV parameters.\ The emission rates become smaller as $\alpha_1$ and $\widetilde{\kappa}_{tr}$ do.

\begin{figure}[!ht]
	\centering
	\includegraphics[width=.48\textwidth]{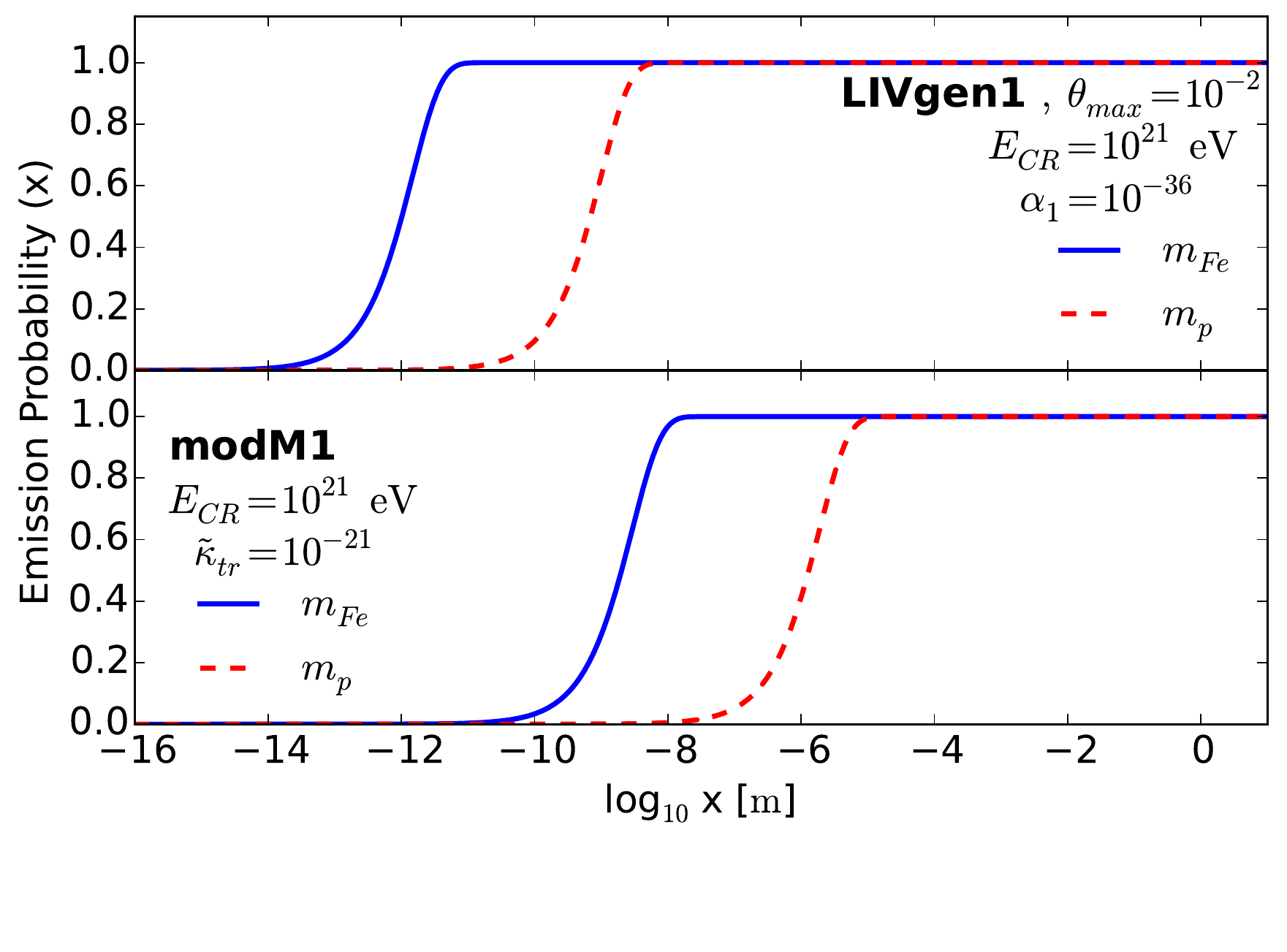}
	\caption{Vacuum Cherenkov emission probability from modified Maxwell theory and LIV generic approach at first order correction.}\label{plot_VCR-P}
\end{figure}

In Fig. \ref{plot_VCR-RE-LIV} it is shown the emitted energy from the generic approach. 
A strong emission rate will present a stronger attenuation effect. 
For an energy scale where vacuum Cherenkov radiation is allowed for the entire mass spectrum, between proton and iron limits, heavier particles will suffer vacuum Cherenkov attenuation some orders of magnitude higher than lighter particles. 
Then, heavier particles will be a\-tte\-nua\-ted sooner in energy. This special behavior is clarified in Fig. \ref{plot_VCR-P} with the emission probabi\-lity as a function of mean free path.
Moreover, in LIVgen1 and modM there are 
regions where vacuum Cherenkov radiation only affects the lighter particles but not heavier, until the energy is high enough that vacuum Cherenkov radiation can affect all the massive particles. 
Therefore, a reduction to the CR flux could be expected due to vacuum Cherenkov radiation that starts with the lighter components. Hence, a tendency to heavy components could be expected in the UHECR flux located between both the thresholds, followed by a reduction on the contributions from the entire mass spectrum at even higher energies. 
From the results in Figures \ref{plot_VCR-massLIV}, \ref{plot_VCR-RE-LIV} and \ref{plot_VCR-P} and for a value for $\alpha_1 = 10^{-36} eV^{-1}$, one could expect that this trend would show up around an energy range of about $10^{18}$ to $10^{19} eV$. Observatories such as Pierre Auger, Telescope Array and HiRes have measured and reported the spectrum of ultra-energy cosmic rays at EeV energies \cite{Auger-Mix,AugerTA-Mix,HiRes-Mix}. Thus,  a dedicated analysis with their data could probe such an effect or at least set limits to $\alpha_n$.

\section{Photon Decay}\label{PD}

\begin{figure}[!ht]
	\centering
	\includegraphics[width=.2\textwidth]{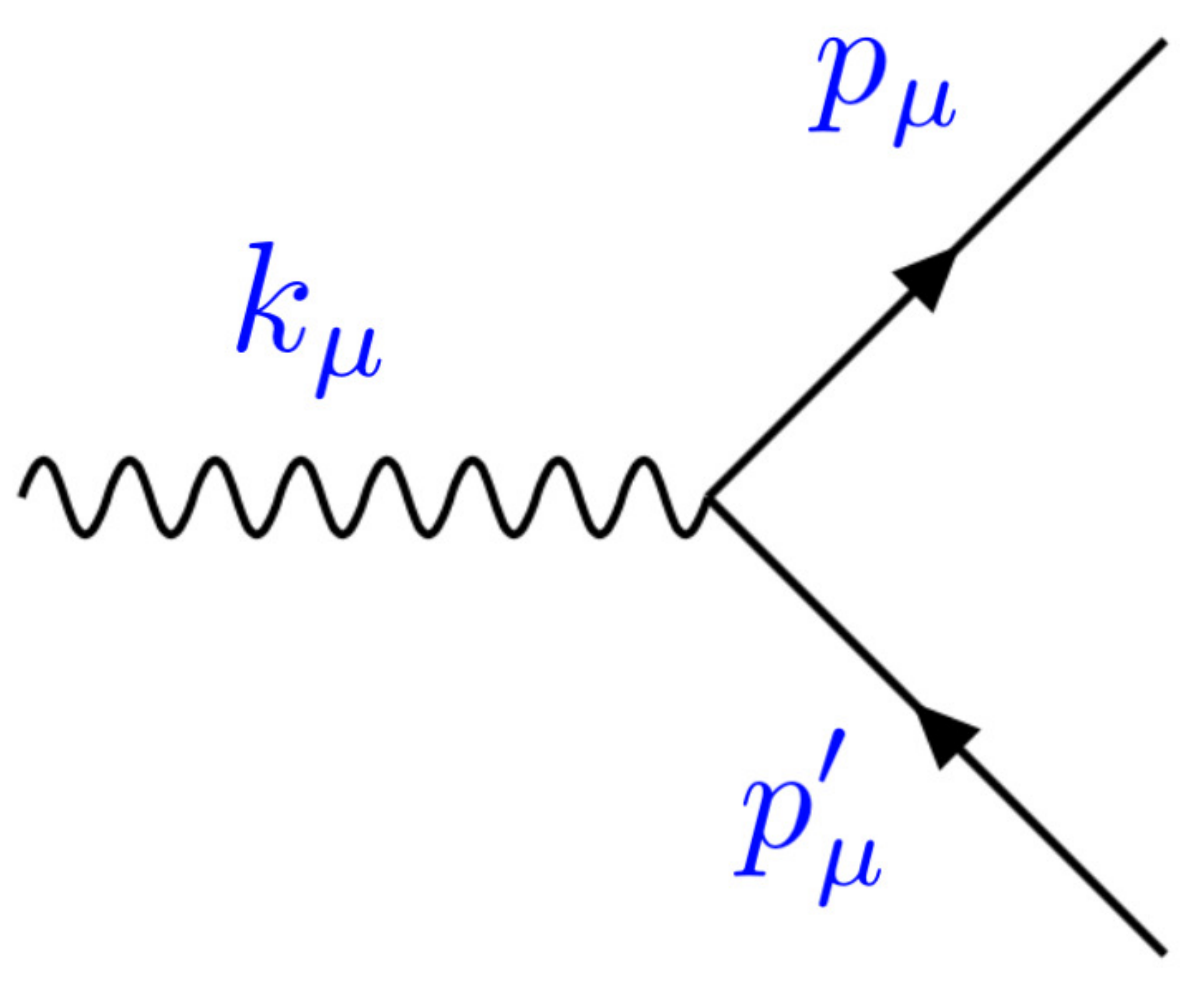}
	\caption{Feynman diagram for LIV photon decay to a pair of leptons. }\label{vertex2}
\end{figure}

Similarly to the process of spontaneous photon emi\-ssion, photon decay, as depicted in  Fig.~\ref{vertex2}, will naturally arise.  This special LIV process has been studied before in the context of generic LIV corrections and the SME (see for instance~\cite{GLASHOW_97,GLASHOW,CROSS,TWO-SIDE,JACOB,GUNTER-PD,GUNTER-PH}). In the present study, we derive the LIV photon decay fo\-llo\-wing the generic construction presented in section~\ref{VCR}, that preserves the generalities of the particular models presented in section \ref{LIV}. 
This turns out to be a very  restrictive scenario that occurs only for  very high energy photons~\cite{GLASHOW_97}. 
Therefore,  it shall be useful to  derive tight upper limits for the generic LIV coefficient in the photon sector.

As we previously did for vacuum Cherenkov radiation, we start the present analysis by considering the LIV photon dispersion relation in  Eq.~(\ref{eq_omega}), $\omega^2  =  k^2 ( 1 + \alpha_n k^n )$.
The decay rate of the process will be thus given by
\begin{equation}\label{eq_deltaPD}
	\begin{aligned}
	d\Gamma = & \dfrac{s}{2} \dfrac{1}{\omega(k,n,\alpha_n)} |\mathcal{M}|^2 \dfrac{d^3 p}{(2\pi)^3 2 E(p)} \dfrac{d^3 p'}{(2\pi)^3 2 E' (p')} \\
	& \times (2\pi)^4\delta^{(4)}(k-p-p')\Theta(p),
	\end{aligned}
\end{equation}
where $(E,\mathbf{p})$ and $(E',\mathbf{p}')$ are the four-momenta components of the final leptons, $l^+$ and $l^-$, respectively. The squared probability amplitude is then given by  Eq.~(\ref{eq_M}) from the previous section. 
Next, we shall perform the integration over $p'$ and write the delta function as follows,
{\small
\begin{equation}
	\begin{aligned}
	 & \delta^{(0)}\left(g(k,p,m,n,\alpha,\sin \theta)\right)  = 
	 \delta^{(0)}\left ( \ \sqrt{1+\alpha_nk^n}~k  \right. \\ 
	& -  \sqrt{p^2+m^2} 
	  \left. -\sqrt{p^2+k^2+m^2-2pk\cos\theta} \ \right)~.
	\end{aligned}
\end{equation}
}

As before, we need to use the $g$ function roots, now  given by the expression,
\begin{equation}\label{eq_PD_g}
	\begin{aligned}
 	(\alpha_n k^n+2\sin^2\theta)p^2 & - 2\alpha_n k^{n+1}\cos\theta p  \\
	& + 2m^2 + \alpha_nk^nm^2=0.
	\end{aligned}
\end{equation}
Notice that unlike to the expressions obtained for the previous process, in here Eq.~(\ref{eq_PD_g}) is a se\-cond order polynomial function in $p$.
To solve it, the condition $k\neq0$ is assumed.  It is not difficult then to see that $p$ will satisfy energy-momentum conservation when,
\begin{equation}\label{disc}
{\small
	\begin{aligned}
	p_{\pm}= 
	 & \left( \alpha_n k^{n+1}\cos\theta  \right. \\ 
	 &\left.  \pm \sqrt{\alpha_n^2 k^{2n+2}\cos^2\theta-4(\sin^2\theta+\alpha_n k^n) (1+\alpha_nk^n)m^2} \right ) \\
	 & \times (2\alpha_nk^n + 2\sin^2\theta)^{-1}. 
	\end{aligned} 
}
\end{equation}

\begin{figure}[!ht]
	\centering
	\includegraphics[width=.48\textwidth]{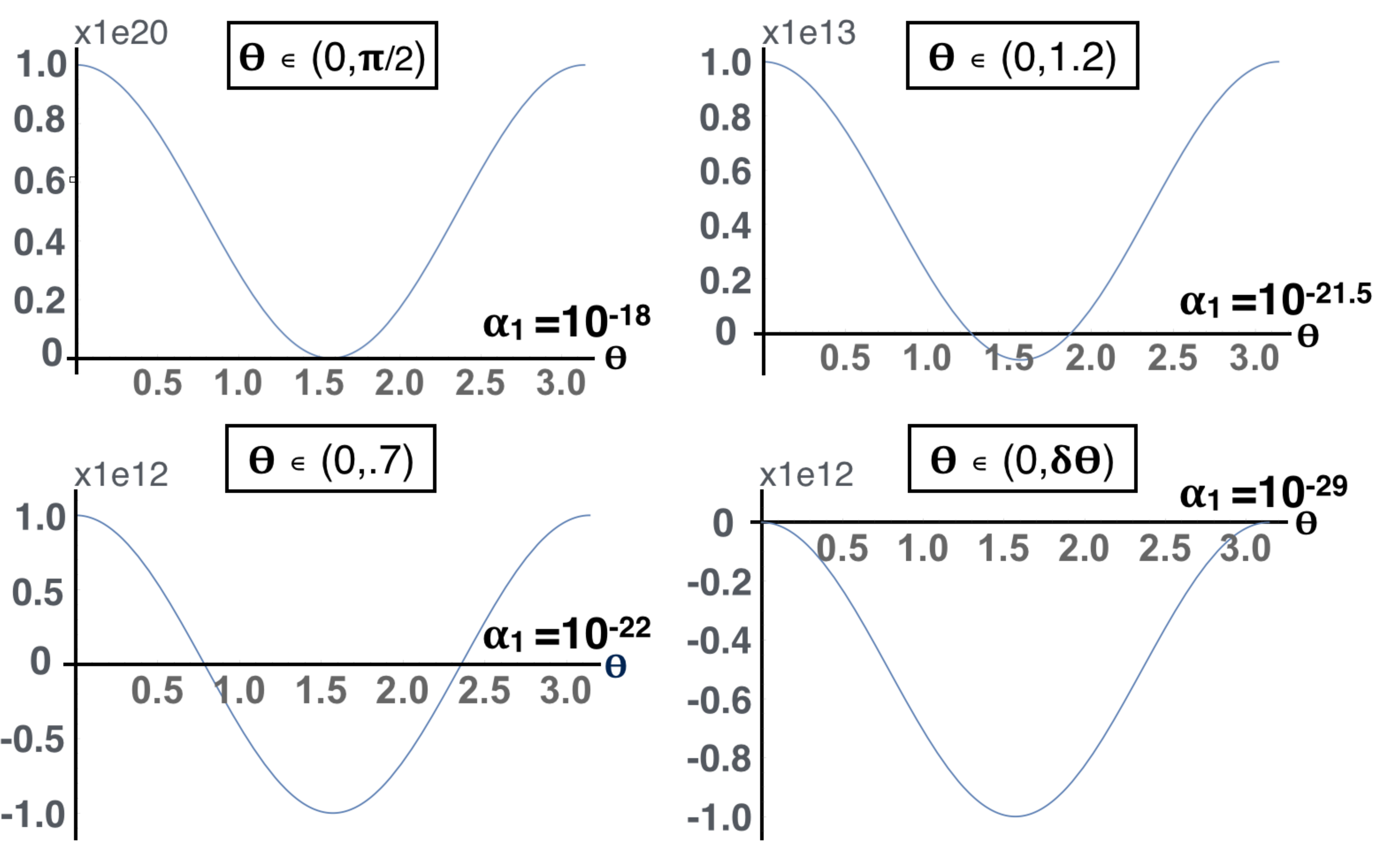}
	\caption{The discriminant behavior from Eq.~(\ref{disc}) for $k=100$ TeV, $m=m_e$, $n=1$ and different values of $\alpha_{1}$. The positive region set the integration limits for the decay rate in Eq.~(\ref{eq_PD_gamma}). }\label{img_PD_det}
\end{figure}

In order to ask for $p$ to be real and positive, the discriminant in Eq. (\ref{disc}) must be real and positive too. 
This condition will restrict the possible emission angles for any given momentum of the photon, with regard to LIV parameters.
To exemplify this restriction, we have plotted the discriminant behavior for a photon energy of $100$ TeV, for $m=m_e$,  $n = 1$  and different values of $ \alpha_1$. The result is depicted in Fig.~\ref{img_PD_det}.
Due to the angular dependence of the function, 
while $\alpha_1 $ decreases, the discriminant becomes negative for a given interval in $\theta$ and the permitted integration region decreases.    
At the limit when $\theta$ goes to zero, the angular integration space becomes null, and the process forbidden. Therefore, this limit establish a threshold on $\alpha_n$ above which LIV photon decay is enable,
\begin{equation}\label{eq_PD_th}
	\begin{aligned}
	\alpha_{n} \ge \frac{4 m^2 }{k^{n}(k^2-4 m^2)}.
	\end{aligned} 
\end{equation}

The analytic expression for the photon decay rate $ \Gamma $ is
\begin{equation}\label{eq_PD_gamma}
	\begin{aligned}
	\Gamma =  &  \dfrac{e^2}{4\pi}  \frac{|4m^2 - \alpha_n k^{n+2}| }{4 k\sqrt{1+\alpha_n k^n}} \\
	& \times  \int_0^{\theta_{max}} \sum_{p_\pm} \frac{p^2\sin\theta d\theta }{| pE' + (p-k\cos\theta)E | }, 
	\end{aligned}
\end{equation}
where
\begin{equation}
{\small
	E=\sqrt{p^2+m^2}, \ \ \ \ E' = \sqrt{k^2 + p^2+m^2-2kp\cos\theta},
}
\end{equation}

\begin{figure}[!ht]
{\centering
	\includegraphics[width=.49\textwidth]{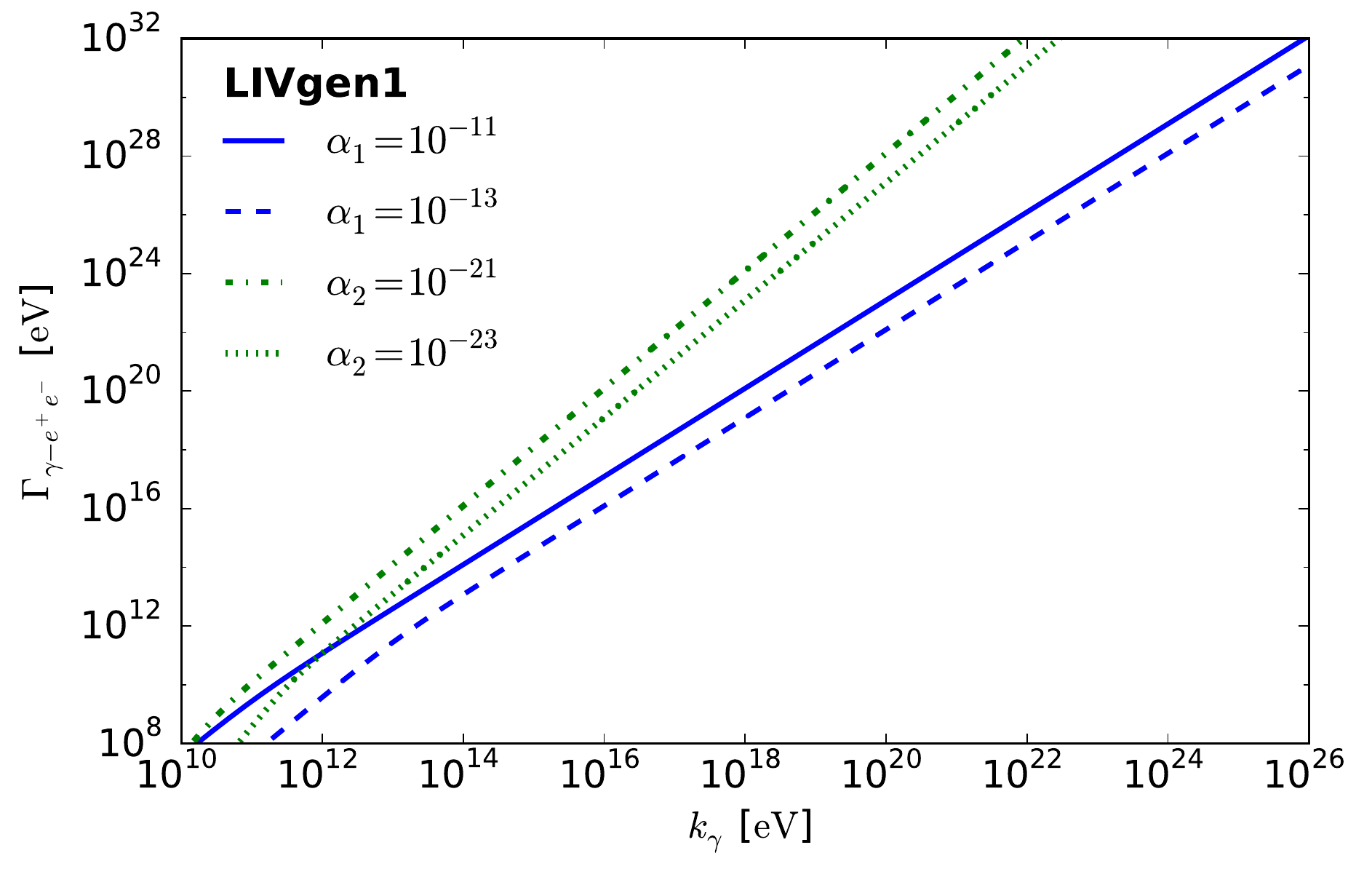}
	\caption{Photon decay rates from LIV generic approach for $n=1,2$ and $\theta_{max} \approx \pi/2$.}\label{img_PD1}
}
\end{figure}

Numerically solving the integral for $n = 1, 2$ one can found the decay rate that can be seen in Fig.~\ref{img_PD1}. It can be appreciated that the decay rates are steadily growing for both $n=1,2$.  They show a change in the slope at an energy inversely proportional to the LIV coefficient $\alpha_n$ and the decay rates become smaller as $\alpha_n$ does. 
 We have found that the case n=0 have a similar behaviour than for $n=1,2$ \cite{Proc3}. In addition, Eqs. (\ref{eq_PD_th}), (\ref{eq_PD_gamma}), for n=0, are compatible, under the assumption that LIV corrections are made only in the photon sector, with the threshold and the decay rate reported in Ref.~\cite{GLASHOW}.

\begin{figure}[!ht]
	\centering
	\includegraphics[width=.46\textwidth]{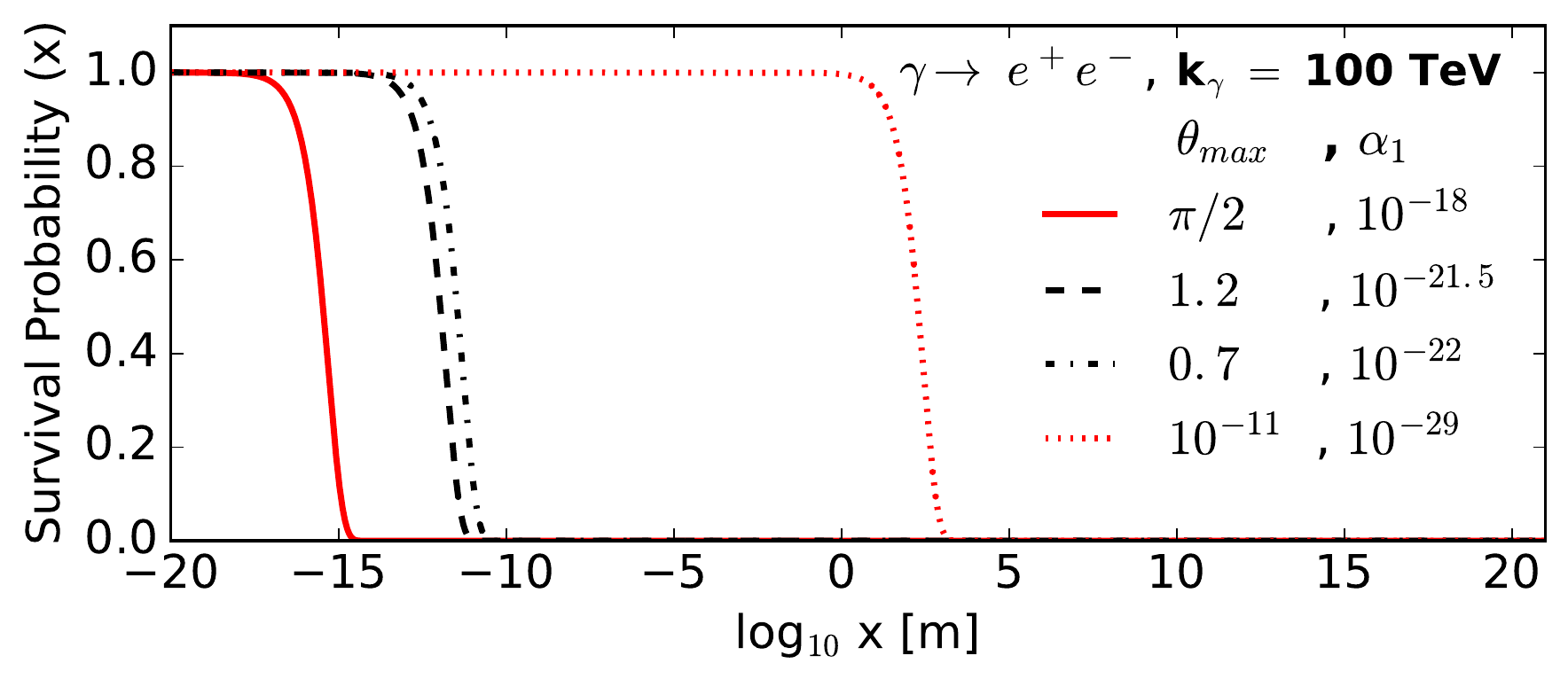}
	\caption{Photon decay probability as a function of the mean free path, for a 100 TeV photon and for the integration limits in Fig. \ref{img_PD_det}.}\label{img_PD2}
\end{figure}

As an example, to illustrate how restrictive this process is, in Fig. \ref{img_PD2} we show the photon survival probability as a function of the mean free path for photons. There we have used the same values we used in Fig.~\ref{img_PD_det}. From this results, it is clear that a 100 TeV photon will not be able to propagate beyond $10^{-12}$~m, for an $\alpha_1=10^{-18}$. However,  in the limit, where $\theta_{max} \to 0$, a photon of 100 TeV will be able to reach us from  kpc distances, but a smaller $\alpha_n$ is implied. 
This would set a critical value that fixes an energy dependent threshold for the LIV parameter. For  any 
$\alpha_n$ above threshold, photon decay becomes extremely efficient. Below  threshold, on the other hand, 
the process is forbidden.
That means, that a lower limit for 
$E^{(n)}_{QG} \approx 1/ \alpha_n^{1/n}$ 
in the photon sector will directly emerge from any observed high energy cosmic photon event, 
of momentum $k_{obs}$, expressed as 
\begin{equation}\label{EQGth}
    E^{(n)}_{QG} > k_{obs}\left[\frac{k_{obs}^2-4 m^2}{4 m^2}\right]^{1/n}.
\end{equation}  
\begin{figure}[!ht]
	\centering
	\includegraphics[width=.49\textwidth]{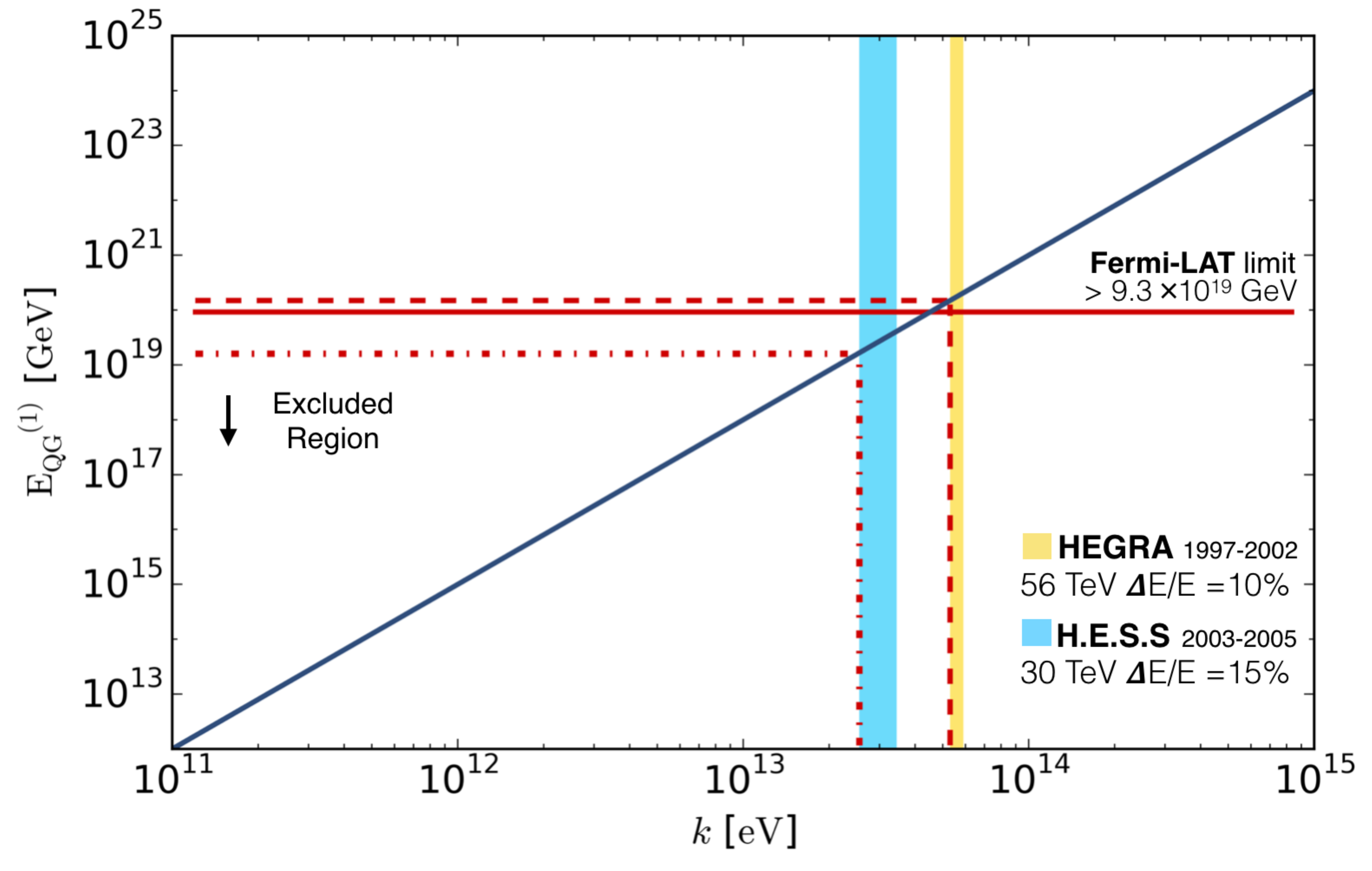}
	\caption{ $E_{QG}^{(1)}$ excluded region from LIV photon decay into electron positron pairs with HEGRA \cite{HEGRA,HEGRA-LIV} and H.E.S.S. \cite{HESS, TWO-SIDE} photon energy measurements. Using Eq. (\ref{eq_PD_th}) we found that $E_{QG}^{(1)}\ge1.5\times10^{20}GeV$ from HEGRA and  $E_{QG}^{(1)}\ge1.7\times10^{19}GeV$ from H.E.S.S., in the $n=1$ case.}\label{img_PD_limit1}
\end{figure}
\begin{figure}[!ht]
	\centering
	\includegraphics[width=.49\textwidth]{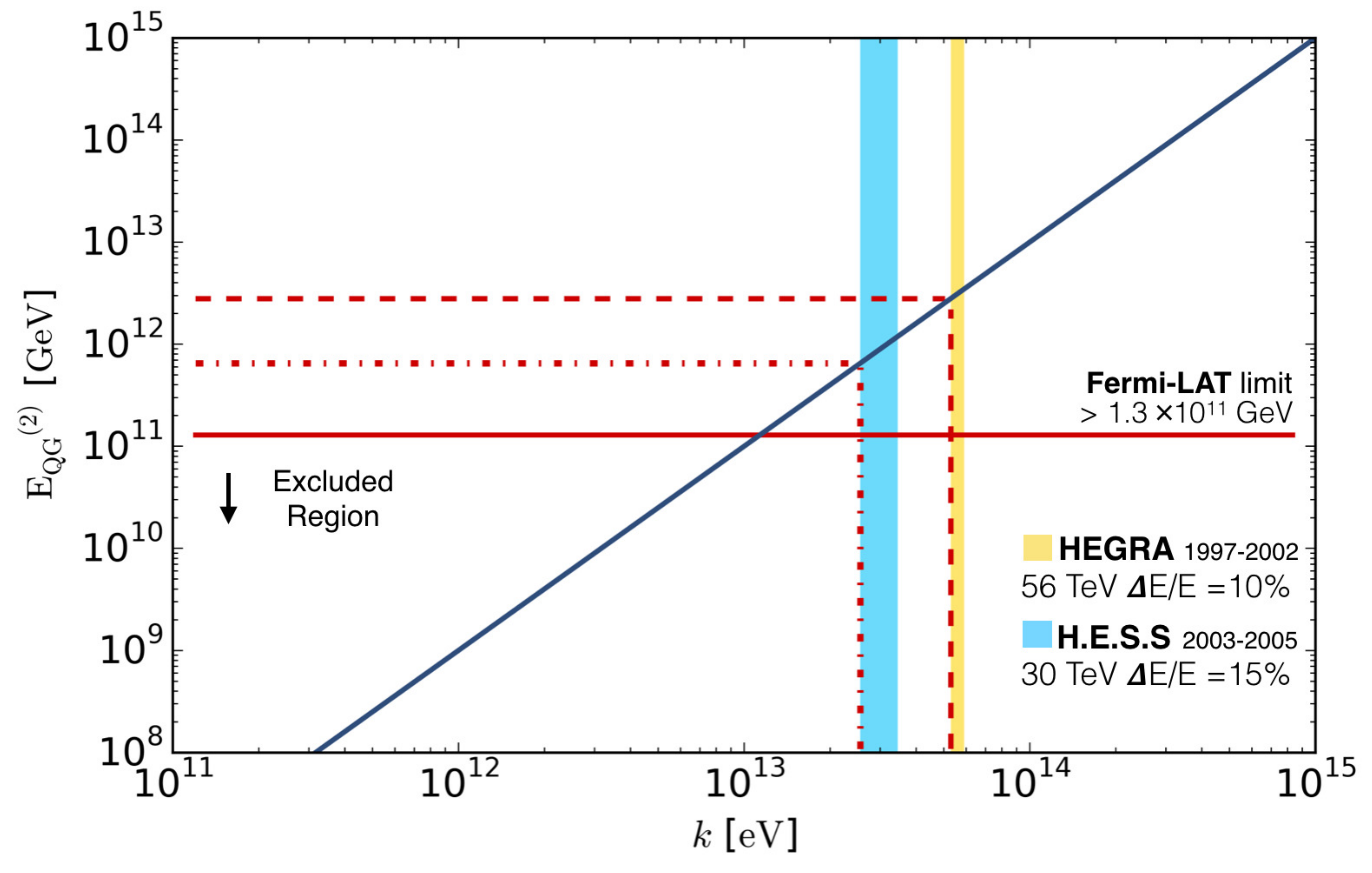}
	\caption{$E_{QG}^{(2)}$ exclude region from LIV photon decay into electron positron pairs with HEGRA \cite{HEGRA,HEGRA-LIV} and H.E.S.S. \cite{HESS,TWO-SIDE} photon energy measurements. Using Eq. (\ref{eq_PD_th}) we found that $E_{QG}^{(2)}\ge2.8\times10^{12}GeV$ from HEGRA and  $E_{QG}^{(2)}\ge6.5\times10^{11}GeV$ from H.E.S.S., in the $n=2$ case.}\label{img_PD_limit2}
\end{figure}

Following this line of thought, and using above expression for photon decay into electron positron pairs, $m=m_e$,
we can depict the  $E^{(n)}_{QG}$ limit as a function of the photon energy event, as it can be seen in Fig.~\ref{img_PD_limit1}, for $n=1$ and Fig.~\ref{img_PD_limit2}  for  $n=2$, represented by the (blue) diagonal line on the plots.
The (red) continuous and horizontal line in both figures is one of the best limits for the LIV generic photon sector proposed by {\it Fermi}-LAT \cite{Example_n}.
The dashed (red) lines are our derived limits from  H.E.S.S. and HEGRA reported data {\cite{TWO-SIDE,HESS,HEGRA,HEGRA-LIV}}. The (colored) bars are the uncertainty of the highest photon energy reported. We are using the lower value as an observed photon energy from astrophysical distances.  In our analysis the best limits are obtained from HEGRA data, as indicated in the figures.  We found that $E_{QG}^{(1)}\ge1.5\times10^{20}GeV$ and 
$E_{QG}^{(2)}\ge2.8\times10^{12}GeV$,
for $n=1$ and $n=2$, respectively.

\section{Cosmic ray photo production}\label{REC}

Another known scenario derived from the generic mo\-di\-fi\-ca\-tion to particle dispersion relations is the one where the thresholds for known processes are shifted to di\-ffe\-rent energies than the ones expected under a LI regime.
However, as we discuss in what follows, there is also an interesting feature derived from this approach that we believe deserves some further attention. It appears in the study of particle production by inelastic collisions, particularly those of the type $AB\rightarrow CD$, in which, in a LI regime, massive final particle states, $C$ and $D$, require some minimal energy from the initial particles $A$ and $B$ for the process to happen. As we will show next, in the generic LIV setup a  higher energy threshold  appears where particle production becomes again forbidden. In the scenario where the collision describes the interaction of an energetic particle pro\-pa\-gating in the cosmic photon background, this last threshold means that at some high energy the cosmic particle shall again freely propagate, at least free from this particular process. This is what we will call a recovery scenario.

For the propose of establishing the above scenario in the analysis below we will apply, in the most general way to all involved particles,  the modification to their dispersion relation, as  presented in Eq.~(\ref{eq_dispersion2}).
To be specific, we have chosen two representative and relevant processes for cosmic ray physics: pair photo production  and pion photo production~\cite{CMBAngelis, GZK-Z,GZK-GK, GZK},
\begin{equation}\label{pair}
	\gamma_{CR} \ \ \gamma_{b} \longrightarrow e^+ e^-,
\end{equation}
\begin{equation}\label{pion}
	p_{CR} \ \ \gamma_{b} \longrightarrow p \ \ \pi^{0},
\end{equation}
where $b$ and $CR$ denote background and cosmic ray primary particles, respectively. 
Notice that we have chosen  protons for the second process, but the analysis would be equally valid for neutrons too.
Unlike to the work presented in the previous sections, here we will not refer to any particular model to support  the main LIV corrections, so keeping the analysis genuinely generic. 
Similar approaches can be found in Refs.~\cite{GUNTER-PH,Stecker2009, Stecker2009NJ}.
Furthermore,  we shall consider a first-order correction function $\Xi$ to  energy-momentum conservation in order to explore the relevance at cosmic ray energies. This correction is motivated by the work in Ref.~\cite{LIV-EM} (for an alternative model see also~\cite{JLuis}), where se\-ve\-ral LIV coefficients were added  in order to properly correct energy-momenta.
To the best of our knowledge such a general approach has not been explored in this general context elsewhere before.
For instance, for pair photo production, in  Eq. (\ref{pair}), let $(\omega,\omega_{b})$ and $(E_+,E_-)$ be the corresponding initial and final energies of particles in the process.  Therefore, in the LIV setup,  we write energy conservation relation as
\begin{equation}\label{eq_Xi}
	E_+ + E_- = \omega + \omega_b +\Xi.
\end{equation}
Or equivalently,
\begin{equation}
	E_+= K (\omega + \omega_b+\Xi),
\end{equation}
and
\begin{equation}
	E_-=(1-K)(\omega + \omega_b+\Xi),
\end{equation}
where $K$ is the inelasticity function of the process. 

Next, we will use the system where the collision is head on; since the  very high and ultra high energy cosmic photons have higher energy than the photon background, $\omega_b \ll \omega$, the final momenta are collinear. Additionally, we will introduce the ultra relativistic regime approximation, that is $m\ll \{ E,p\} \ll M$ 
and we will keep first order LIV coefficients only, since they are expected to be small.   

For pair photo production, let the invariants be
 \begin{equation}\label{eq_invPP1}
	S_{photon}= \omega^2 - k^2 = -\alpha_n \omega^{n+2},
\end{equation}
and 
\begin{equation}\label{eq_invPP2}
	S_{\pm}= E^2_{\pm} - p_{\pm}^2 = m_e^2 - \alpha_{\pm,n} E_\pm^{n+2},
\end{equation}
where we have inserted a different LIV parameter for the electron sector to account for a possible particle dependence. 
Previous work have studied similar corrections but considering a momenta correction with the opposite sign in $\alpha_n$~\cite{GUNTER-PH}.
Substituting and equating  the initial and final invariant to first approximation in LIV coefficients, we found that
\begin{equation}\label{eq-PP-1L}
{\small
\begin{aligned}
4&\omega  \omega_b  - m_e^2 \left( \frac{1}{K(1-K) }
-\frac{m_e^2}{2K(1-K)(\omega + \omega_b +\Xi)^2} \right)\\
&= \alpha_n\omega^{n+2}\left[ 1 + \frac{\omega_b^{n+2}}{\omega^{n+2}} - \frac{\omega_b}{\omega} 
\left( 1+ \frac{\omega_b^{n}}{\omega^{n}}\right) \right] \\
&+\alpha_{+,n}K^{n+1}(\omega + \omega_b)^{n+2}\\
&\ \ \ \ \times \left[ -1 + \frac{m_e^2}{2} \frac{1}{(1-K)(\omega + \omega_b)^2}\right] \\
&+\alpha_{-,n}(1-K)^{n+1}(\omega + \omega_b)^{n+2} \\
& \ \ \ \ \times \left[-1 + \frac{m_e^2}{2} \frac{1}{K(\omega + \omega_b)^2}\right]~.
\end{aligned}
} 
\end{equation}
Note that $\Xi$, the LIV contributions to energy-momenta conservation, was reduce to one term.
By demanding that $m_e,\ \omega_b, \ \Xi \ \ll\omega$,  the last expression becomes, at the lower correction order,
\begin{equation}\label{eq-PP-1}
4\omega  \omega_b  - \frac{m_e^2}{K(1-K)} = \beta_n \omega^{n+2}~,
\end{equation}
where we have defined 
$$\beta_n =  \alpha_n - \alpha_{+,n}K^{n+1} - \alpha_{-,n}(1-K)^{n+1}~.$$
Notice that the contribution of $\Xi$ has disappeared. 
This clearly shows that, at leading order, LI momentum conservation rule effectively holds for high energy cosmic particle collisions.

In the limit where $\alpha_n=\alpha_{\pm,n}=0$, the RHS of Eq.~(\ref{eq-PP-1}) vanishes, thus providing a unique solution to $\omega$ in terms of $K$, given by $\omega_0 = \frac{m_e^2}{4K(1-K)\omega_b}$.
For $K=1/2$, the standard LI pair production threshold for a cosmic photon will arrive as usual, where
\begin{equation}\label{eq_LIPP}
	\omega_{th}^{LI}=\frac{m_e^2}{\omega_b}~.
\end{equation}
Otherwise, Eq.~(\ref{eq-PP-1}) is a polynomial equation of order $n+2$ on $\omega$, which in general may have more than one real solution. This shall be the core of our argument below.
It is instructive to rewrite Eq.~(\ref{eq-PP-1}) in a more simple and generic form by defining the dimensionless variables 
\begin{equation}
	x_\gamma = \frac{\omega}{\omega_0}\quad \text{and}\quad
	\Lambda_{n,\gamma}= \frac{\omega_0^{n+1}}{4\omega_b} \beta_n ~.
\end{equation}
In such terms the polynomial threshold equation for pair photo production becomes
$$\Lambda_{n,\gamma} x_\gamma^{n+2} - x_\gamma +1 = 0~.$$
Last equation is consistent with the result presented in  Ref.~\cite{GUNTER-PH} for pair production under a generic LIV correction and assuming LI energy-momentum conservation.

Before analysing the implications of last equation, lets move towards the exploration of the general photo production process $A\omega_b \rightarrow CD$, to determine its corresponding threshold equation. Particular cases will be pair and pion photo production by cosmic nucleons.
Following a similar path as before, we would start by writing down  the corresponding modified dispersion relations 
for the involved particles. Aside to Eq. (\ref{eq_invPP1}) for the background photon, we would also have
$ E_{i}^2 - p_{i}^2 = -\alpha_{i,n}E_{i}^{n+2}$, for $i={A,C,D}$, respectively.
A similar algebra as before would then give the following equation for the threshold, at the lower order, for high energy cosmic ray primaries,
\begin{equation}\label{pithreshold}
	4\omega_bE_A +m_A^2 -\frac{Km_C^2 +(1-K)m_D^2}{K(1-K)} =  \Upsilon_n E_A^{n+2}~,
\end{equation}
where, assuming that $\omega_b\ll E_{A}$, 
\begin{equation}
  \Upsilon_n  =   \alpha_{A,n}  - \alpha_{D,n}K^{n+1} - \alpha_{C,n}(1-K)^{n+1}~.
\end{equation}

Note that, if we set all LIV coefficients to zero, Eq.~(\ref{pithreshold}) implies the solution
\begin{equation}
E_{A0} = \frac{Km_C^2 +(1-K)m_D^2}{ 4\omega_b K(1-K)}-\frac{m_A^2}{4\omega_b}~.
\end{equation}
Clearly, pair photo production formulae is recovered if one takes $m_A=0$, and $m_{C,D}=m_e$.
By taking  $m_{A,C}=m_{p}$ for a proton, $m_D=m_\pi$ for pion and $K=\frac{m_\pi}{m_\pi + m_{p}}$, we get the known LI threshold for neutral pion photoproduction,
\begin{equation}\label{eq_LIPi}
	E_{p_0}  = \frac{m_\pi^2 + 2m_\pi m_p}{4\omega_b}~.
\end{equation} 
This particular threshold plays a relevant role in UHECR since the end of CR flux is commonly associated to the GZK cutoff \cite{GZK-GK,GZK-Z,GZK}.

Once more, by introducing dimensionless variables in the general case, now given  by
\begin{equation}
x_A=\frac{E_A}{E_{A0}} \quad \text{and} \quad 
\Lambda_{n,A}=  \frac{E_{A0}^{n+1}}{4\omega_b} \Upsilon_n~,
\end{equation}
we  finally get the same form for the threshold equation as in the pair photo production case. Therefore, irrespective of the process we can express such an equation as
\begin{equation}\label{lambda}
	\Lambda_n x^{n+2} - x +1 = 0,
\end{equation}
where in general $x=E^{LIV}/E^{IL}_{th}$ and $\Lambda_n$ contains an energy independent linear function, $\{\beta_n,\Upsilon_n\}$, on all LIV parameters. 
Clearly, if $\{\beta_n,\Upsilon_n\}=0$, LI regime would be recovered.

It is worth stressing that in both the cases, Eq.~(\ref{lambda}) is, to first approximation in LIV terms, independent from particle nature, the functional form of A in Eq.~(\ref{eq_dispersion2}) and $\Xi$ in Eq.~(\ref{eq_Xi}). Last means that the results do not depend on lineal corrections to the energy-momentum laws. 

\subsection{The LIV secondary threshold}

\begin{figure}[h]
	\begin{center} 
  		\includegraphics[width=.44\textwidth]{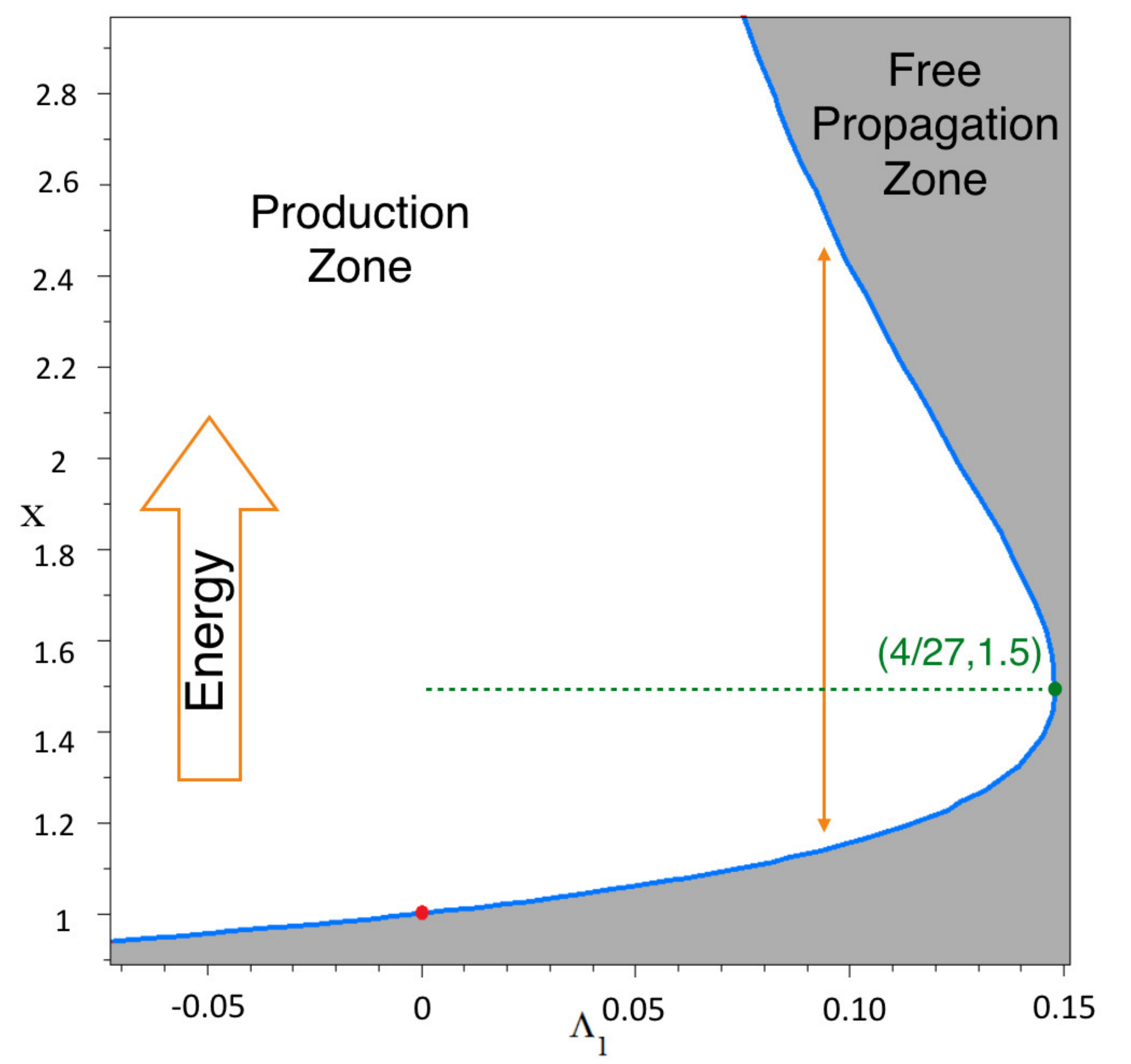}
	\end{center}
	\caption{\small{ Solutions to the threshold equation for $n=1$, as a function of $\Lambda_1$. The $f(x)$ roots separate  free propagation and photo production zones as indicated.  Critical point (in green) indicates  the upper limit on $\Lambda_1$. LI case is stressed by a (red) point at $\Lambda_1=0$.}}\label{fig_recovery}\hfill
\end{figure}

Let us  next consider the implications of LIV  threshold equation (\ref{lambda}). Clearly, to each value of $\Lambda_n$ it shall correspond a different scenario which describes the behavior of photo production processes. 
For simplicity we will first consider in detail the lower order correction, $n=1$, and then comment on the general case. In this case we get the cubic equation
$f(x) = \Lambda_1 x^3 - x+1= 0$, which in general has three different solutions, but for the LI case at $\Lambda_1=0$. As it is easy to check, the possible roots of $f(x)$ provide three different physical scenarios,  as follows:

\begin{itemize}
    \item[i)]  For $\Lambda_1<0$ there is only one real root for $f(x)$. Other two roots  are complex, and thus, non phy\-sical. Furthermore, the real root turns out to be always positive but smaller than one.  This implies that the derived LIV physics  produces a threshold at lower energies  than that expected in the LI regime 
    ($E^{LIV}_{th} < E^{LI}_{th}$). This means an earlier activation of photo production in energy than expected. 
    The gap among the two thresholds will increase as $\Lambda_1$ goes to more negative values.     
    
    Note that for an universal $\alpha_1$ we get $\beta_1=\Upsilon_1= \alpha_1 K(2-K)$, and thus  
    this scenario only arises provided that $\alpha_1<0$.
    
    \item[ii)] There is a critical value at $\Lambda_1=4/27$ above which $f(x)$ has only one negative and two complex roots. Therefore, for $\Lambda_1> 4/27$, there is no physical solution to the threshold equation. As a consequence,  photo production process becomes forbidden and the Universe gets transparent to the propagation of the cosmic ray primary. Such an scenario is of course in plain contradiction with astrophysical observations and thus, one must impose the 
    consistency condition
    \begin{equation}\label{eq_reality}
	\Lambda_1 < 4/27.
    \end{equation} 
    Note that such a critical value do exist for any arbitrary  $n$. A simple algebra  shows that real roots demand that $\Lambda_n<\Lambda_{nc}=(n+1)^{n+1}/(n+2)^{n+2}$, which is a quickly decreasing function on $n$. 
    At critical value, the only root is $x_{nc}= (n+2)/(n+1)$. 

    \item[iii)] Finally, for $0<\Lambda_1<4/27$ the threshold equation has three real solutions. Nevertheless, only two of them are positive, and physically acceptable. A non trivial but important consequence arises here. A range of $x$ values shall now become bounded by two thresholds.  
    
\end{itemize}

The whole effect of these scenarios can easily be understood if we plot $f(x)$ roots as a function of $\Lambda_1$, as we have done in Fig.~\ref{fig_recovery},  and notice that the co\-rres\-pon\-ding curve actually divides the parameter space into two regions. Clearly, by simply looking at the particular solution where LI holds (for $\Lambda_1 =0$), we can identify the free propagation zone below the curve (where $E<E_{th}$). This zone is analytically connected to the region on the RHS of the plot, where $\Lambda_1>4/27$, in accordance to our previous discussion. Furthermore, it is connected to the region above the curve, where  $0<\Lambda_1<4/27$. 
The conclusion is unavoidable. Whereas the lower root gives the expected energy  threshold where photo production gets switched on, the second root, at higher energies,  represents a new energy threshold where photo production shall, once more, stop.

This conclusions are consistent with the findings in Ref.\cite{GUNTER-PH} for pair photo production, although it is worth stressing that this physics is rather common to all photo production processes. Furthermore, it is not difficult to realize that there always exist a secondary threshold for any value of $n$, and for $0<\Lambda_n<\Lambda_{nc}$. As a matter of fact, RHS of Eq.~(\ref{lambda}) has a linear behaviour for small $x$, with a negative slope, whereas, for large $x$, it is dominated by the single term $\Lambda_n x^{n+2}$, with a positive derivative. Accordingly, there exist a single absolute minimum, at $x_{min} = [(n+2)\Lambda_n]^{-1/{n+1}}$, that lays in the energy region where photo production is on, beyond the first root (above the first threshold). 
This, in turn,  gives rise to a second root, and therefore, to the secondary threshold. That would be the signature of the generic LIV for any photo production process.

This phenomena would produce an {\it opacity band} for a propagating primary.  Out of the band ultra high energy cosmic ray particles, like protons and very high energy photons, would be free of the photo production processes produced by the interactions with background photons. 
Additionally, the bandwidth decreases for larger values of $\Lambda_n$. Whereas the first threshold can only take values from $ x \in[1,x_{nc}) $,  the second threshold decreases very fast as $\Lambda_n$ grows, until the photo production processes fade out,  when $\Lambda_n$ reaches its critical limit, $\Lambda_{nc}$. 

The appearance of an opacity band should have visible effects in the integrated cosmic ray flux. Although a detailed calculation of this is out of the scope of the present work, our results suggest that potentially visible effects are possible. In general grounds, for individual particles, one would expect that for cosmic ray primaries with energies within the band, the physics should be very similar to the standard LI scenario. Photo production processes will contribute to  quickly diminish the energy of the proton primary, and to the disappearance of high energy primary  photons. However, if energies well above the second threshold were accessible to the primaries, they will see a transparent Universe along its propagation, which means no depletion on their flux would be expected.
In this sense the measured flux would show a recovery behavior at the largest energies.

As an  example, for  $\Upsilon_1 \geq 1.67 \times 10 ^{-45} \ eV^{-1}$, primary cosmic protons would avoid  pion photo production, and hence the GZK threshold \cite{GZK-GK,GZK-Z,GZK}. 
From the present analysis if $\alpha_{p,1}=\alpha_{\pi,1}=0$, then $\alpha_{\gamma,1} \gtrsim -10^{-20} \ eV^{-1}$ will be strong enough to produce a noticeable threshold change below $10^{20} \ eV$. However, further observations of UHE cosmic and gamma rays are essential. LIV studies under the Colleman and Glashow approach, through the correction of inelasticity,  that contrasts with the experimental data of Pierre Auger Observatory and HiRes can be found in Refs.~\cite{Stecker2009,Stecker2009NJ}.

A potential smoothing of the effect is possible, once the specifics of flux injection and energy distribution of the photon background be taken into account in the calculation of the expected cosmic ray fluxes. But we believe this  deserves some further consideration, since it may serve to probe the limits for the LIV generic approach.

\section{Conclusions}

Generic LIV, expressed as the deformation of particle dispersion relations, by energy/momentum dependent functions, can be used to study the possible  LIV physics, without relaying on specific models, and thus, in a more direct and simple manner. 
As a way to demonstrate this point, we have addressed three different physical processes that are relevant in the study of the propagation of the very to ultra high energy cosmic rays: vacuum Cherenkov radiation, photon decay, and photo production processes. Whereas the first two are considered as characteristic of LIV, since they are forbidden in LI schemes, photo production turned out to also provide interesting signatures for LIV effects.

First, Cherenkov radiation analysis has been used to validate the generic approximation. 
To this end, we have elaborated a calculation for the emission rate, which takes into account the standard amplitudes derived from QED rules, but explicitly incorporating the LIV modification of the photon dispersion relation, to order $n$ in the photon momentum. All LIV effects for fermions were neglected, though, in order to be able to compare the results to those of models based on the spontaneous breaking of Lorentz Invariance.
The analysis shows that the  emission rate does present the same qualitative behaviour
as  the outcomes obtained from the modified Maxwell Theory and the Maxwell-Chern-Simons Theory:  
emission  rate grows  with the charged particle energy
and become smaller as $\alpha_n$ does,
but it presents a drop ending threshold which is  inversely dependant on the LIV coefficient. This threshold sets the energy value below where  LI physics is recovered.  
We have also analysed the mass and charge composition sensitivity of the emission rate, 
since this results would have some impact on the analysis of the attenuation process along the propagation of high energy cosmic ray primaries. Our results show a clear dependency on the cosmic primary mass and charge, having a stronger emission rate for heavier nuclei, but with a similar slope and a higher energy threshold. Similar results are found when the modified Maxwell Theory is used for the calculations. 
This would imply that for energies  below the LIV Cherennkov thresholds, cosmic ray flux should be consistent with LI physics. As we move up, just passing the proton emission threshold, protons will be attenuated along propagation, and the heavier mass cosmic ray component would become dominant. As energy passes the threshold for heavier masses, the cosmic rays flux should move again towards the lighter component mass, as 
a stronger attenuation of heavier particles appears. A detailed calculation of this, that take into consideration the astrophysical models for particle injection, is needed to better understand the extension of the effect for the ultra high energy cosmic rays.

The implementation of the generic approach to photon decay on vacuum 
proves that  this process, if allowed,  is very efficient and stringent for the propagation of the cosmic ray primary.  
Nevertheless, as we have argue,
phase space for outgoing fermions vanishes when photon energy is below a critical threshold energy, defined through the expression $E_{QG}^{(n)}>k[(k^2 - 4m^2)/4m^2]^{1/n}$. At such energies, the total decay rate becomes zero and the process prohibited, so recovering the standard LI results.  Above this threshold, on the other hand, photon decay rate quickly becomes very large. Thus, the process strongly restricts the possible propagation of the photon to very short distances from source, to the extent that it becomes unlikely for a primary photon to travel astronomical distances. 
The aftermath of this is a novel, direct and very simple way to bound LIV energy scale, $E_{QG}^{(n)}$, through the direct observation of the very high energy photons. Using reported gamma telescopes data and photon decay into electron positron pairs, we have established, for $n=1$, the stringent bound for the LIV scale at $E_{QG}>1.5\times 10^{20}~GeV$, coming from HEGRA most energetic events.

Finally, generic LIV effects incorporated to all particles involved in the most general  
photo production process, where a high energy cosmic particle interacts with the photon background, producing 
two particles at final state, had been shown to modify the LI production threshold. 
Interestingly enough, our analysis shows that LIV effects are completely general, at first approximation. They do not depend on the particular nature of the particles involved in the process, neither depend on possible lineal corrections to energy momentum conservation. Regardless to the specific photo production process,  LIV  shifts the production threshold towards lower or higher energies. This solely depends on the sign of the overall LIV parametric function, $\Lambda_n$, that encodes LIV effects and which enters in a polynomial like term of order $n+1$ that  corrects the, formerly linear, threshold equation.
Furthermore, as we have argued, this changes imply an upper bound for the LIV parameter, at $\Lambda_n>(n+1)^{n+1}/(n+2)^{n+2}$. For larger values of $\Lambda_n$ photo production remains forbidden. Consequently,
the shifting on the LIV production threshold is predicted to always be smaller than a factor of $(n+2)/(n+1)$ for $\Lambda_n>0$.
Even more importantly, our analysis has uncovered a novel signature for the LIV physics of photo production processes, that is worth to underline.  Derived from the LIV modification to the universal threshold equation, a secondary threshold at higher energies will always appear, for  $\Lambda_n>0$, where particle production becomes again forbidden. Above such threshold,  cosmic particles shall again propagate freely, giving rise to a possible recovery on the cosmic ray flux.

Physical implications of the purely LIV secondary threshold are interesting. A cosmic particle primary shall encounter an
opaque Universe along its propagation, due to photon background,  only for primary energies within the two thresholds. At such energies, photo production will attenuate the energy of the particle, moving it to lower values where the process is switched off, permitting the further free propagation. Of course this is in general the standard LI scenario, but for the shifting of the actual position of the threshold. On the other hand,  if particle primaries were injected by astrophysical sources at energies well above the LIV secondary threshold, they would not suffer attenuation due to photo production. A recovery on the cosmic particle primary flux would be expected as a consequence above the opacity band defined by the thresholds. 

Nevertheless, a word of caution to our above naive observations is in order. Even though our conclusions are genuine and had to be considered valid for any local cosmic ray particle photo production process, the actual calculation of the effects that LIV corrections, and second threshold, should have on the observed cosmic ray flux had, yet, to be calculated with care.  Actual calculation has to take into account a number of issues. A complete analysis of the processes including arbitrary angle emission has to be considered, as well as the details of the injected primary flux and the photon background energy distribution.
Besides, second threshold happens at the very high energies. Few astrophysical sources are known to produce cosmic ray primaries at such regimes. Even if they do so, candidates are always far enough as to make cosmic expansion effects to be relevant. Redshift diminishes by itself the energy of propagating particles, which means that even if a charged particle  primary is produced at high enough energies at source reference frame, local interactions with photon background, on our observer reference frame, would be happening at lower energies. On our frame, at some critical distance between source and observer, the cosmic primary might reach energies where the local Universe becomes again opaque, and energy attenuation would take place as usual. This may still have, however, distinctive features.  It shall effectively move the second threshold effects towards an apparently higher energy on source basis, whereas it would move the first threshold closer to the LI value. 
But it could also modify the injected flux in a distance to source dependant way, that, if enough data were available, could be distinguishable on the features of the observed flux.    
Comparison among a set of well known sources, located at different redshifts, may serve as a probe to establish observational bounds for the LIV parameters. We believe this possibility deserves further study. 
\bigskip

\section*{Acknowledgments}
This work was partially supported by Conacyt Mexico, under grant 237004.

\bibliography{bibfile}

\end{document}